# Perspective

# High entropy van der Waals materials

Tianping Ying[1,2,†], Tongxu Yu[3,†], Yanpeng Qi[4], Xiaolong Chen[1,*], Hideo Hosono[2,*]

[1.] Beijing National Laboratory for Condensed Matter Physics, Institute of Physics, Chinese Academy of Sciences, Beijing 100190, China

[2] Materials Research Center for Element Strategy, Tokyo Institute of Technology, Yokohama 226-8503, Japan

[3.] Gusu Laboratory of Materials, Jiangsu 215123, China

[4.] School of Physical Science and Technology, ShanghaiTech University, 393 Middle Huaxia Road, Shanghai 201210, China

**Abstract**

By breaking the restrictions on traditional alloying strategy, the high entropy concept has promoted the exploration of the central area of phase space, thus broadening the horizon of alloy exploitation. This review highlights the marriage of the high entropy concept and van der Waals systems to form a new family of materials category, namely the high entropy van der Waals materials (HEX, HE = high entropy, X= anion clusters) and describe the current issues and next challenges. The design strategy for HEX has integrated the local feature (e.g., composition, spin, and valence states) of structural units in high entropy materials and the holistic degrees of freedom (e.g., stacking, twisting, and intercalating species) in van der Waals materials, and has been successfully employed for the discovery of high entropy dichalcogenides, phosphorus tri-chalcogenides, halogens, and MXene. The rich combination and random distribution of the multiple metallic constituents on the nearly-regular 2D lattice give rise to a flexible platform to study the correlation features behind a range of selected physical properties, e.g., superconductivity, magnetism, and metal-insulator transition. The deliberate design of structural units and their stacking configuration can also create novel catalysts to enhance their performance in a bunch of chemical reactions.

## 1. Introduction

Just like mathematicians said that some infinities are bigger than other infinities in number fields, there are some disorders that are more complex (and intricate) than other disorders in the physical world. Physicists coined the term 'entropy' to describe the distributions of myriad states from the universe to the atomic nucleus under thermodynamic constraints. In the area of materials science or condensed matter physics, the large ensemble of atomic constituents (~$10^{23}$ $cm^{-3}$) lends a broad playground for the concept of entropy and disorder to implement their roles and exhibit intriguing features which are intractable, e.g., the Kauzmann entropy paradox [1] in glass transition. We still have no rigorous theoretical tools to elucidate the correlations between various aspects (i.e., structural hierarchy, dynamical heterogeneity, configuration frustration, thermodynamic fluctuation, etc.) of the vitreous states with holistic disorders. Compared with this vexing situation, the distribution disorder with a spatial regularity is somewhat within reach of the existing theoretical framework, e.g., Parisi's replica symmetry solution of spin glass[2–4].

High entropy alloy (HEA), theoretically proposed in the 1980s and experimentally realized in 2004, equips materials scientists and physicists with a proper distribution disorder to expand their toolkit[5–8]. The impetus of fusing more principal elements into one single alloy system is to explore the uncharted central area of phase diagrams to achieve more excellent mechanical performance[9–11]. The high entropy concept implies the disordered distribution and imperceptible correlation of different constituents. Since then, the advantage and prospect of HEA have propelled this alloying strategy to a range of material categories, e.g., oxides[12], nitrides[13], carbides[14], borides[15], etc[16]. Up until now, both HEA and these high entropy ceramics are bulk systems with isotropic nature. That means they can be employed as structural materials under severe conditions. However, for some specific usage such as lubrication and catalysis, the bulk form makes a large part of the materials used ineffectively. Under these situations, the explicit need for a large surface/volume ratio and effective slip plane requires a family of materials with anisotropic structural features.

With this consideration, the possibility of imbuing the high entropy concept into the flatland of van der Waals materials (vdW materials) has been exploited by our group[17,18] and a few others[19–23]. The strong intralayer covalent interaction and weak interlayer interaction make the vdW materials easy to be exfoliated. Coincidently, the discovery of graphene is in the same year as the experimental realization of HEA[24]. Likewise, the fascinating behavior of graphene excites the research community to uncover a broad range of physical systems with monolayer or few-layer nature, e.g., phosphorene, borophene, MXene, and two-dimensional transition metal dichalcogenides (2D TMD)[25,26]. Both the HEA and vdW materials have rich tunable degrees of freedom and corresponding bountiful properties[27,28]. The marriage of the concepts of HEA and vdW materials offers us a platform to look forward to the large possibilities of high entropy van der Waals materials (HEX).

The myriad literature on vdW materials and HEA systems makes it difficult if not impossible to encompass all the exciting developments in a single review article. Compared with that, the nascent state of HEX research permits us to cover most of the recent work within our reach. We hope this review of HEX entwining the two concepts of HEA and vdW materials will intrigue the interest of the broad research community to explore the fascinating landscape of this new flatland.

## 2. General features

Due to the 3D isotropic nature of the HEAs that have been synthesized up until now, the constituent distribution disorder plays the main role in the culprit of high entropy in discussions of these systems. The configurational entropy of the system is typically expressed as $S=-R\Sigma \ln N$, when the principal elements in the system are isomolar, and R is the gas constant and N is the number of different kinds of elements in the alloy. Nevertheless, this definition is unable to reflect the influence of dimensionality and anisotropy on the configurational entropy, thus we coined another definition to describe the difference within high entropy systems with distinct dimensional features. Assume the number of microstates Ω in each spatial dimension of a high entropy system is identical, then $S_{1D} = k_B \ln \Omega$, $S_{2D} = k_B \ln(\Omega \times \Omega) = 2k_B \ln \Omega$, and $S_{3D} = k_B \ln(\Omega \times\Omega \times \Omega) = 3k_B \ln \Omega$ for isotropic 1D, 2D, and 3D systems, respectively. So, there is an entropy deficit of 33% by dimension reduction from 3D to 2D. This entropy deficit can be somewhat compensated by a series of modulation strategies granted by the characteristics of vdW materials, e.g., the distinct stacking order/disorder of multilayers, the random distribution of the intercalated ionic/molecular species, and the intrinsic fluctuation of the two-dimensional layered structures.

### 2.1 Structural units

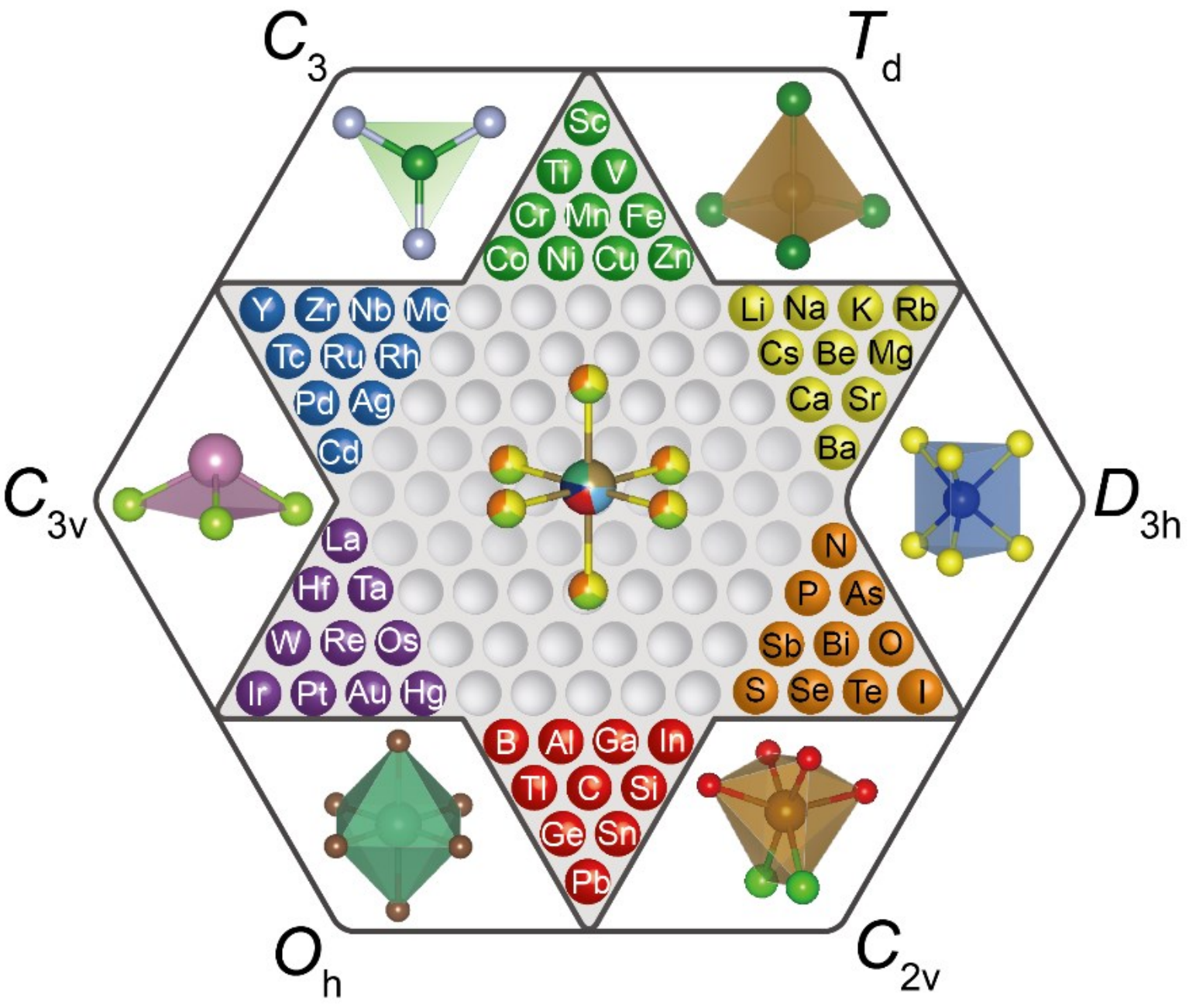

Figure 1. The "Chinese checkers" design strategy of high-entropy vdW materials. The color balls placed at each corner of the hexagram represent the optional multi-elements that can be used to construct the structural units. Several typical structural units are exhibited in the voids of the hexagon circumscribing the hexagram. Their corresponding point symmetry groups are marked at the corner of the hexagon. Inside the checkerboard, a $HES_6$ octahedron is sketched for illustration.

The detailed structural description and characterization of disordered condensed matter systems is always a daunting problem. Even though the high entropy materials more or less retain the long-range lattice structure, the multiple on-site elements and multifarious distribution make the traditional unit cell description unreliable. A more tractable method is to adopt the characteristic structural unit concept, which is widely used in structural chemistry to reflect the microscopic/mesoscopic configuration and employed to delineate the short-range order of the liquid and vitreous states of condensed matter. From this perspective, the two hundred 2D compounds thus far discovered can be disentangled into a handful of building blocks, as shown in Fig. 1 and Table 1. These building blocks generally determine the function and band alignment of the host compound. Starting from these basic structural units, the construction of new functional 2D materials is just like playing with Lego by stacking these building blocks through the corner-, edge- or face-sharing means. Thus far, this description is equivalent to the unit cell method under the umbrella of the symmetry of space groups, which leads to a numbered combination of the materials for specific structural units. Apparently, the most efficient way to break this limitation is to enrich the multiplicity of the building blocks. Introducing the concept of high entropy to the design of these blocks is a tempting option. Indeed, the discovery of HEX materials will extend the investigation of 2D materials from limited combinations to unbounded possibilities.

**Table 1. Structural features of some common building blocks in vdWs materials.**

| Name | Formula | Point group | Example |
|---|---|---|---|
| Triangle | $AB_3$ | $C_3$ | BN, graphene |
| Tetrahedron | $AB_4$ | $T_d$ | FeSe |
| Octahedron | $AB_6$ | $O_h$ | $CdI_2$ |
| Prism | $AB_6$ | $D_{3h}$ | 2H-$MoS_2$ |
| Triangular pyramid | $AB_3$ | $C_{3v}$ | InSe, GaS |
| Gyroelongated square pyramid | $AB_4C_5$ | $C_{4v}$ | PbClF |
| Cubic | $A_6B_{12}C_8$ | $T_d$ | $AuTe_2Se_{4/3}$[29,30] |

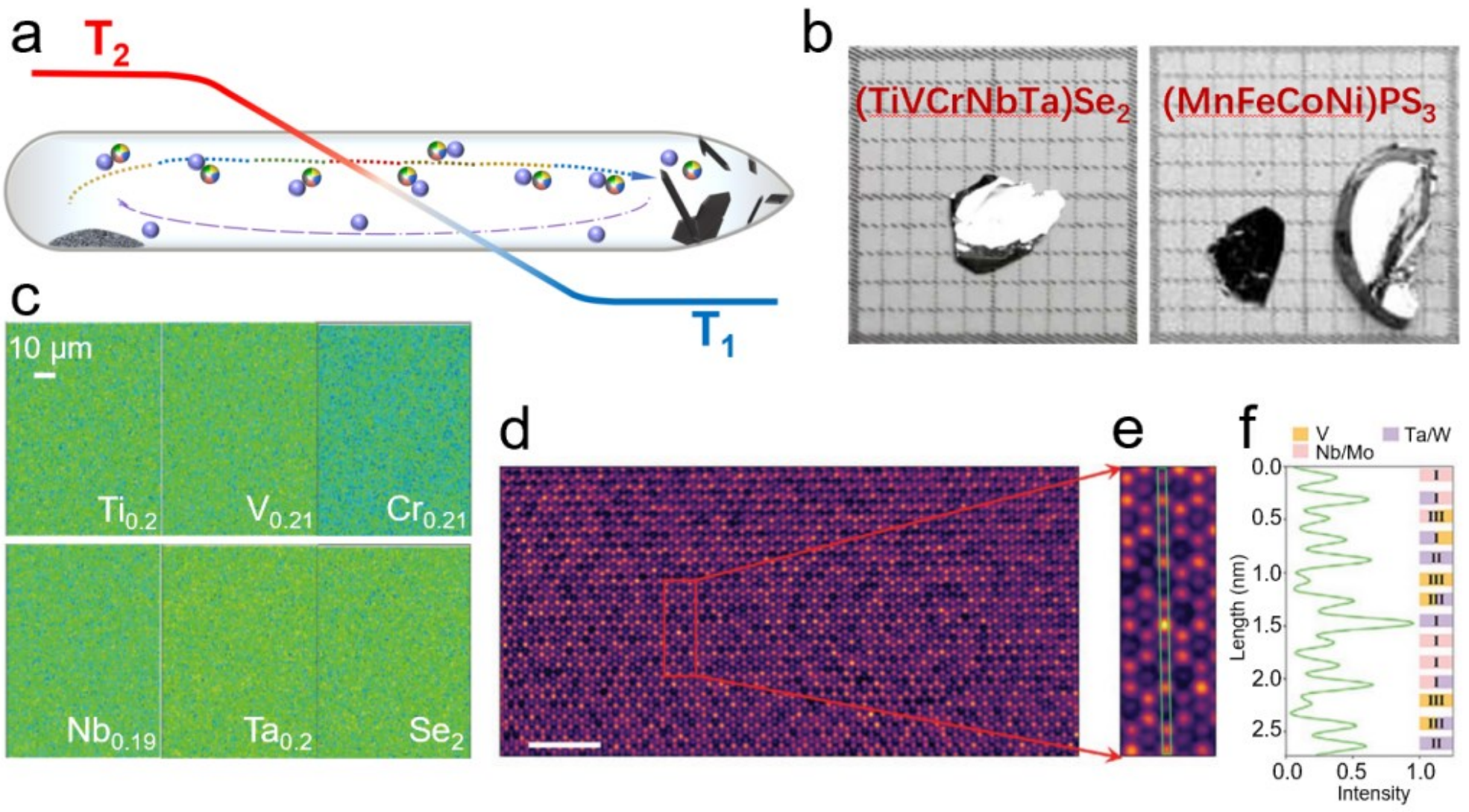

Figure 2. Crystal growth and element distribution of HEX materials[17,19]. (a) Schematic diagram of the chemical vapor transport method suitable for the growth of $HES_2$, $HESe_2$, $HECl_2$, $HEBr_2$, $HEI_2$, and $HEPS_3$. (b) Optical images of single crystals of $(TiVCrNbTa)Se_2$ and $(MnFeCoNi)PS_3$ with sub-centimeter sizes. (c) Quantitative element mapping of $(TiVCrNbTa)Se_2$ by EPMA. (d) Atomic-resolution scanning transmission electron microscopy high-angle annular dark-field (STEM-HAADF) image of a flake of $(MoWVNbTa)S_2$, where the highlighted region (red box) shows the [001] projection. (e) STEM-HAADF image showing local variations in the intensity of atomic columns due to varying composition of cations. (f) The intensity profile of the green box region is illustrated in (e), where the legends correspond to the predominant elements in each atomic column.

## 2.2 Crystal growth and element distribution

To obtain the single phase of high entropy materials is always a main concern for the research community. Phase separation including microphases is unpalatable in the exploration of HEAs. Different from the indispensable harsh synthesis condition to obtain single phase HEAs, the growth of HEX single crystals is relatively easy. This can more or less attribute to the covalent/ionic connection inside and between the structural units of HEX, which restrain the thermodynamic wobbling and jumping of the constituents often unavoidable in HEAs and other vitreous states. Most of the single crystals grow to millimeter size or even several centimeters (Fig. 2a, b) by using the conventional chemical vapor transport method. EDX and EPMA mapping show that all the alloyed elements are homogeneously distributed in the sample (Fig. 2c).

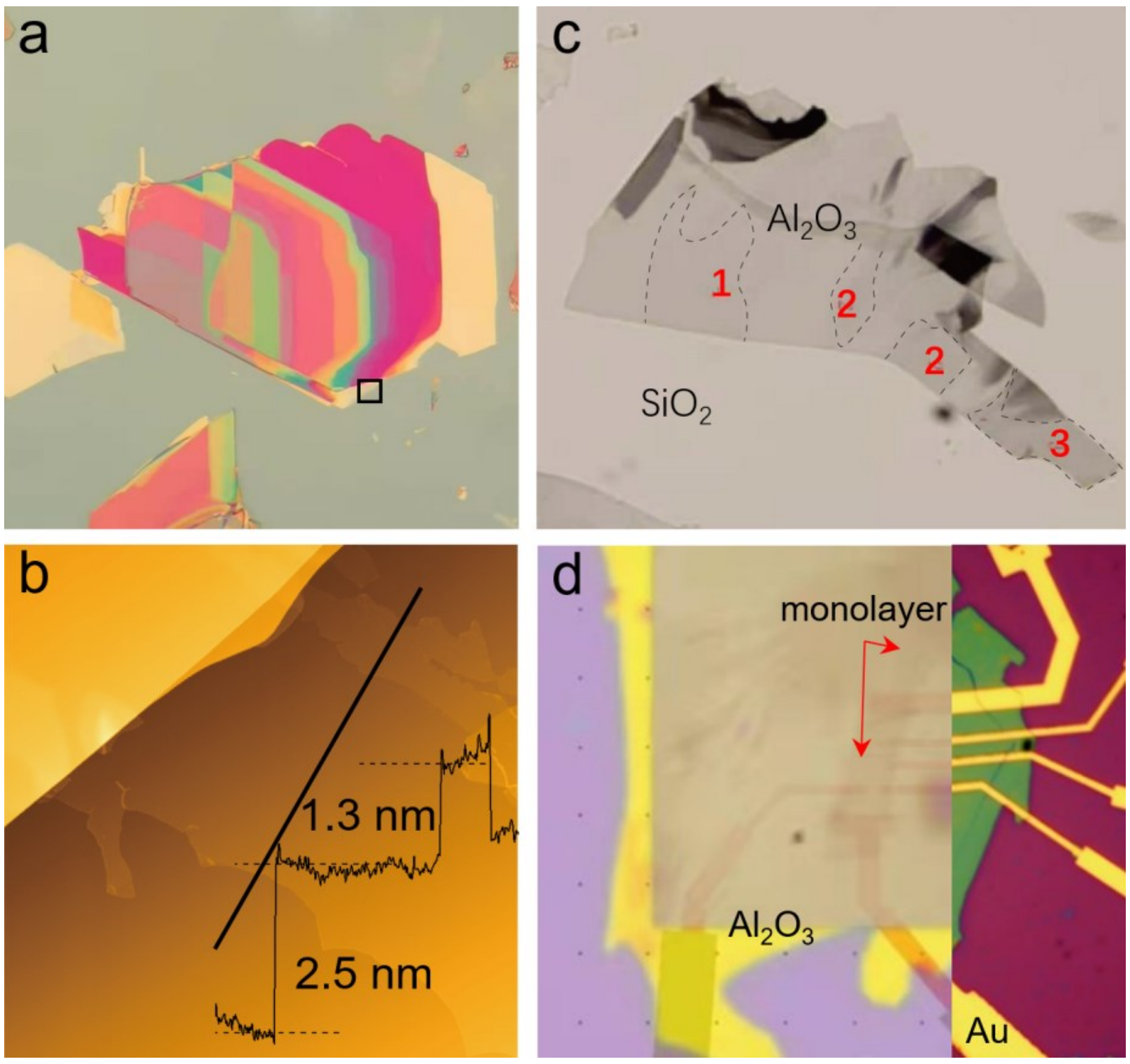

Figure 3. The exfoliation of several typical HEX single crystals[17]. (a) Optical image of the exfoliated (MnFeCoNi)$PS_3$ on a $SiO_2$/Si wafer. (b) Atomic force microscope (AFM) topographic image of the squared area shown in (a). The cross-sectional profile along the black line is superimposed. (c) Optical image of as-cleaved monolayer and few layers (TiVCrNbTa)$S_2$ by using $Al_2O_3$-assisted exfoliation method. Layer numbers have been indicated. (d) A typical device fabricated on monolayer (TiVCrNbTa)$S_2$.

**2.3 Exfoliation and intercalation**

Inherited from its 2D ancestor, the interlayer interactions of HEX are dominated by van der Waals force. Thus, these crystals can be easily exfoliated into a few layers (Fig. 3a, b) by using Scotch tapes. Monolayer can also be acquired using the $Al_2O_3$-assisted exfoliation method (Fig. 3c) recently developed by Y. B. Zhang *et al.*[31]. The bulk crystal was exposed to an oxygen partial pressure of $10^{-4}$ mbar while Al was thermally evaporated to form an $Al_2O_3$ film on it. With the help of the strong adhesion between the bulk crystal and the $Al_2O_3$ film, it is able to exfoliate the crystals to monolayers, which is difficult by traditional techniques. After that, the $Al_2O_3$ film and fragments of HEX microcrystals that had been detached from the bulk were picked up by a thermal release tape for further use. Different from all the reported high entropy materials (including alloy, borides, oxides, carbides, and nitrides), HEX is the first one that can be exfoliated into few layers or even monolayer.

The ability to accommodate interlayer species (atoms, ions, molecules) by intercalation is another unique feature of vdW materials. Figure 4 shows the pristine and the intercalated $K_x(NH_3)_y$(TiVCrNbTa)$S_2$ as an example. The apparent peak shift towards lower angles indicates the c-axis is much enlarged after the co-intercalation of potassium and ammonia by around 2.5 Å per unit cell (Fig. 4a). Further investigation into the intercalation effect on the system reveals a kind of charge accumulation behavior with the element-selective feature. We trace the XPS peak evolution of different elements with the intercalation of monovalent (K) and divalent (Ba) species in (TiVCrNbTa)$S_2$. As shown in Fig. 4b-4d, peak positions of S and Ta monotonically shift to lower binding energy, in line with the continuing charge donation from the intercalated cations, while V and all the rest elements are not sensitive to the intercalation of Ba. The underlying mechanism is not clear so far. An intriguing observation is that by immersing these intercalated compounds in ethanol or water, it is feasible to obtain thin flakes in large quantities. Such few-layer or even monolayer HEX nano-flakes can make full use of the exposed surface area and can be particularly useful for catalysis and energy storage. It is yet unknown how the intercalation may affect their physical properties such as magnetization, superconductivity, and optical properties.

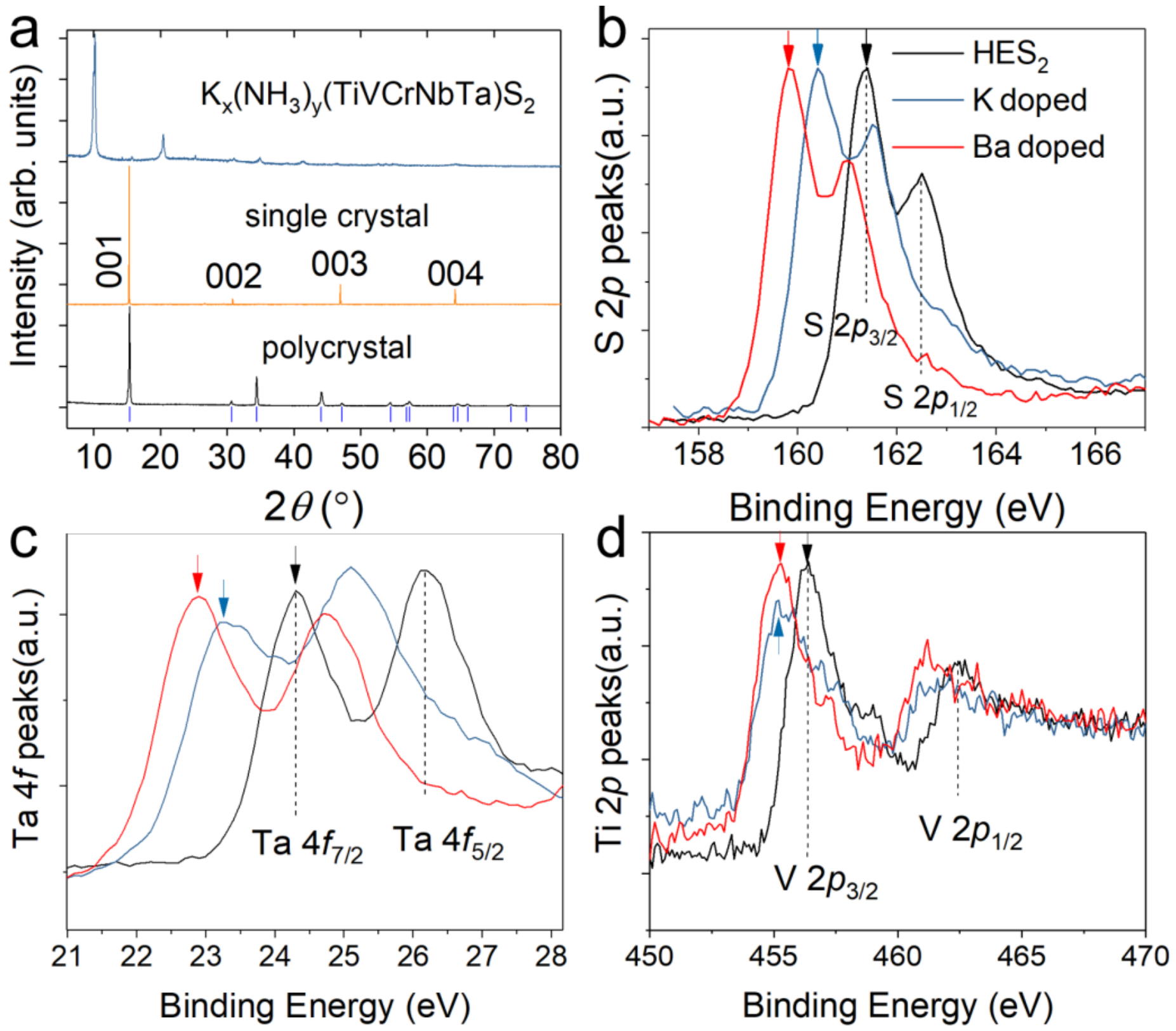

Figure 4. The intercalation of HEX[17]. (a) X-ray diffraction patterns of polycrystalline (black), single-crystalline (orange), and potassium-intercalated (blue) (TiVCrNbTa)$S_2$. (b-d) XPS peaks of S 2*p*, Ta 4*f,* and V 2*p* states for raw, monovalent (K), and divalent (Ba) intercalated (TiVCrNbTa)$S_2$. A continuous redshift by increasing the doping content can be seen in S 2*p* and Ta 4*f*. However, the distinction between the monovalent and divalent doping effect cannot be distinguished in V *2p* (as well as in the rest transition metals), indicating the uneven and element-selective charge distribution within the high-entropy system.

### 2.4 Phonon properties and vibrational modes

The inevitable lattice distortion introduced by the mixture of multiple constituents with different atomic sizes is supposed to have a profound influence on the phonon properties and vibrational modes of high entropy systems. A naïve imagination is that the random distribution and the frustrated magnetic order (if any) will broaden the characteristic peaks of the Raman and infrared spectra. We take $HEPS_3$ as an example because its individual counterparts such as $FePS_3$ and $MnPS_3$ have been well-studied for their magnetic response by Raman spectroscopy and $HEPS_3$ itself has a well-defined antiferromagnetic ordering at 70 K (vide infra). $FePS_3$ is an Ising-type AFM ($T_N$ = 118

K) with a ferromagnetic arrangement along the zigzag chain and an antiferromagnetic arrangement with adjacent chains.

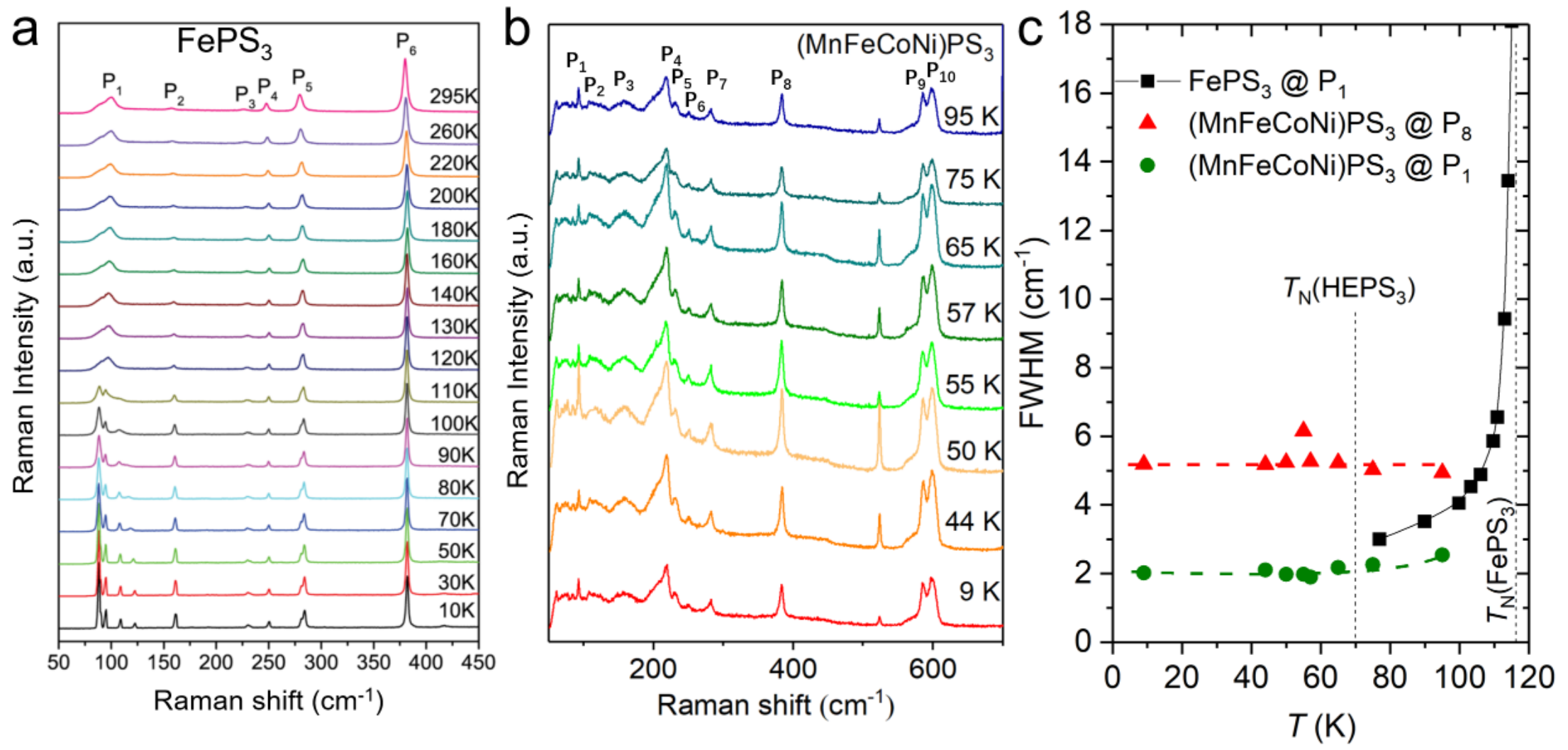

Figure 5. Structure information from Raman spectroscopy. (a) Temperature dependence of Raman spectra of bulk $FePS_3$[32]. (b) Temperature dependence of Raman spectra of $(MnFeCoNi)PS_3$ with an antiferromagnetic transition at 70 K. (c) Plots of full width at half maximum (FWHM) of Raman peak of $P_1$ (88 $cm^{-1}$, 92.7$cm^{-1}$) and $P_8$ (383 $cm^{-1}$) versus temperature in $FePS_3$ and $HEPS_3$. The broken lines indicate the Neel transition temperatures for $FePS_3$ and $HEPS_3$ at 118 K and 70 K, respectively.

Once the temperature is lowered below $T_N$ (Fig. 5a), the Raman active mode of $P_1$ in $FePS_3$ quickly splits into four sharp peaks, indicating the enlargement of the magnetic cell and folding of the modes in other high symmetry points to $\Gamma$. Similar behavior can be found in other mono-transition metal $MPS_3$ compounds, where the noticeable broadening of $P_1$ is closely related to the crossover from AFM ordering to paramagnetic around Neel temperature. Interestingly, $HEPS_3$ shows completely different behavior. As shown in Fig. 5b, all the vibrational modes are clearly distinguishable and relatively strong and remain almost identical from 95 K to 9 K, crossing the AFM transition at 70 K. The $P_1$ peak, which is generally deemed as the indicator of the magnet-lattice ordering, remains quite sharp even above $T_N$. We extracted the full width at half maximum (FWHM) of $P_1$ and $P_8$ to show the trend more clearly in Fig. 5c. Here we give a tentative explanation based on our present results. The FWHM of P1 slowly but steadily increases above 60 K. It is possible that this peak will diverge approaching the highest AFM transition temperature of the element in the alloy, for example, 155 K as in $NiPS_3$. This observation indicates the coexistence of both long-range order ($T_N$ = 70 K in $HEPS_3$) and short-range order, meanwhile, the individual alloyed metals may dictate the behavior of Raman peaks. Such alloying indeed changes the magnetic behavior and drives the system away from Vegard's law.

## 2.5 Corrosion resistance

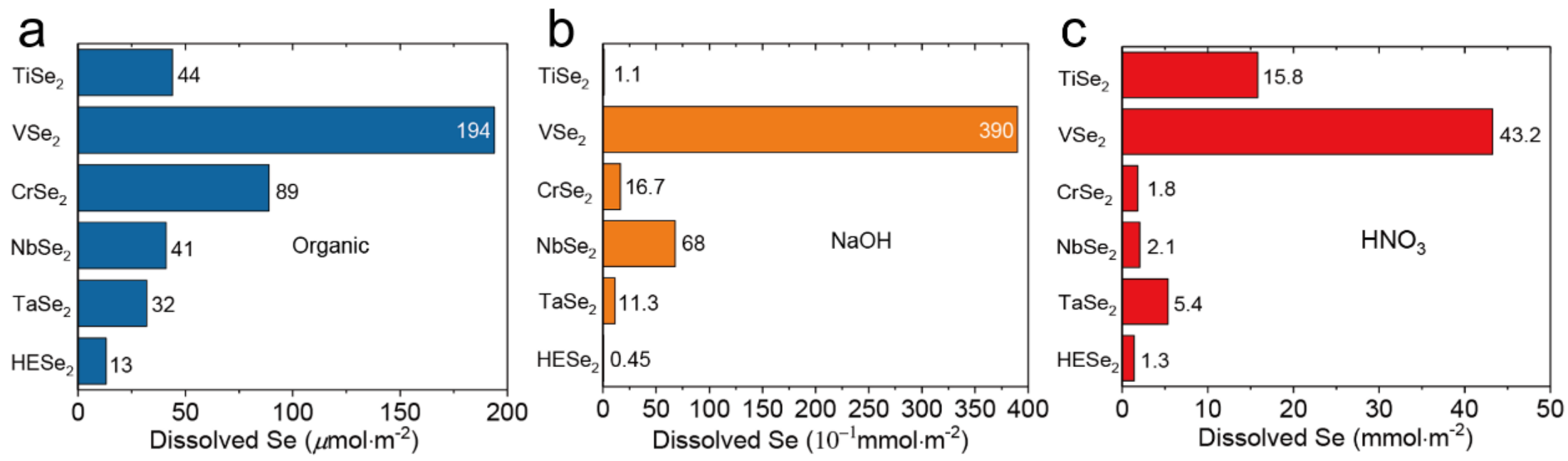

Figure 6. Corrosion resistance of HE-dichalcogenides[17]. (a-c) Corrosion resistance measurements of $(TiVCrNbTa)Se_2$ and its individual components against acid ($HNO_3$), base (NaOH), and organic (butylamine mixed in tetrahydrofuran) reagents.

A broad family of HEAs has been reported to show excellent corrosion resistance against harsh chemical environments and excellent catalysis properties[33–37]. HEX also inherited this feature. Figures 6a-6c show the dissolved Se of $HESe_2$ and its mono-metallic di-selenides in typical acid ($HNO_3$), base (NaOH), and organic reagents. Note that the organic solution and the treatments are based on standard treatment of the synthesis of organic ureas[38]. Similar behavior has been found in $HEPS_3$ and HE-MXene systems. The superior robustness of HEX towards all the rest simple compounds highlights the effective enhancement of chemical stability of the high entropy systems in reduced dimensional systems. The improved corrosion resistance is especially beneficial for practical uses of various heterogeneous catalytic reactions, where the key challenge is to avoid the solvation of the active components (vide infra).

**2.6 Phase Formation**

Even though the covalent/ionic nature of the atomic links broadens the thermodynamic window of single-phase HEX compared with the HEA systems, the local irregular links between the multiple structural units and following holistic distortion of the lattice can still trigger the instability of the single-phase and induce phase separation. Table 2 summarizes the explored HEX materials by our group and the reported literature up to date. The failed element combination to obtain single phases of HEX serves as a valuable resource for us to decipher the formation rule of HEX. Although high-throughput computation has been applied to predict the formation of single-phase HEAs[39], it becomes much more complex when anions in HEX are involved. There are two (out of four) widely used empirical rules for the search of HEA referenced from the well-known Hume-Rothery rules[40,41], where only elemental properties are considered. Namely the requirement to synthesize single phases of HEA asserts that the dispersion of both the electronegativity difference ($\Delta\chi = \sqrt{\sum_{i=1}^{n} c_i(\chi_i - \bar{\chi})^2}$) and the atomic size mismatch ($\delta = 100\sqrt{\sum_{i=1}^{n} c_i(1 - r_i/\bar{r})^2}$) of different systems is constrained. The other two qualitative criteria of similar crystal structures and same valency are less important for HEA as all the metals adopt one out of the three common structures (body-centered cubic, hexagonal close-packed, or cubic close-packed) under thermodynamic equilibrium conditions and have zero valence. Statistical analysis of all the experimental data of HEX systems is summarized in Fig. 7a, from which we could see that the constraint of the former still roughly holds, namely, the single phases (solid symbols) fall into a narrow electronegativity difference range. But the atomic size mismatches are more dispersed. This is because, in HEX, different from HEA dominated by metallic bonds, the metallic ions are coordinated by anions to form plural M-X ionic bonds. Thus, we suggest the feature of the structural units, such as the average bond length of M-X or the average volume of the polytope, rather than the atomic size mismatches, could be a better gauge. We use crystalline 3D visualization software such as VESTA[42] to automatically determine the average bond lengths of the structural units from databases like ICSD. We define the characteristic bond mismatch of M-X as

$$\beta = \sqrt{\sum_{i=1}^{n} c_i \left(1 - \frac{r_{M-X(i)}}{\overline{r}_{M-X}}\right)^2}$$

As shown in Fig. 7b, once using $\beta$ as the abscissa, the initially-jumbled data points are more or less separated, where the single phases are collapsed to the corner with $\Delta\chi$ < 15~20% and $\beta$ < 1.3~2%. This is especially conclusive for larger anions in systems like $HETe_2$, $HEPSe_3$, and $HEI_2$. The $\Delta\chi$-$\beta$ diagram can provide general guidance in searching for more HEX systems. Adopting the volume characteristic of the respective polytope gives a similar result. It is also worth noting that in a certain system, i.e., $HEPS_3$, the $MS_6$ octahedron shows a much larger tolerance towards distortion than in other systems, where the single-phase and the phase-separated ones are still intermingled (Fig. 7c). For example, phase separation occurred in $Mn_{0.6}Fe_{0.2}(CoNi)_{0.2}PS_3$ ($\beta$=2.32%, $\Delta\chi$=15.46%) judging from the diffraction pattern, whereas single crystals can be easily prepared in $(FeMnCoNi)PS_3$ ($\beta$=2.47%, $\Delta\chi$=14.29%). Maybe the detailed elemental misfit cannot be blurred out by the structural unit features in such a specific system.

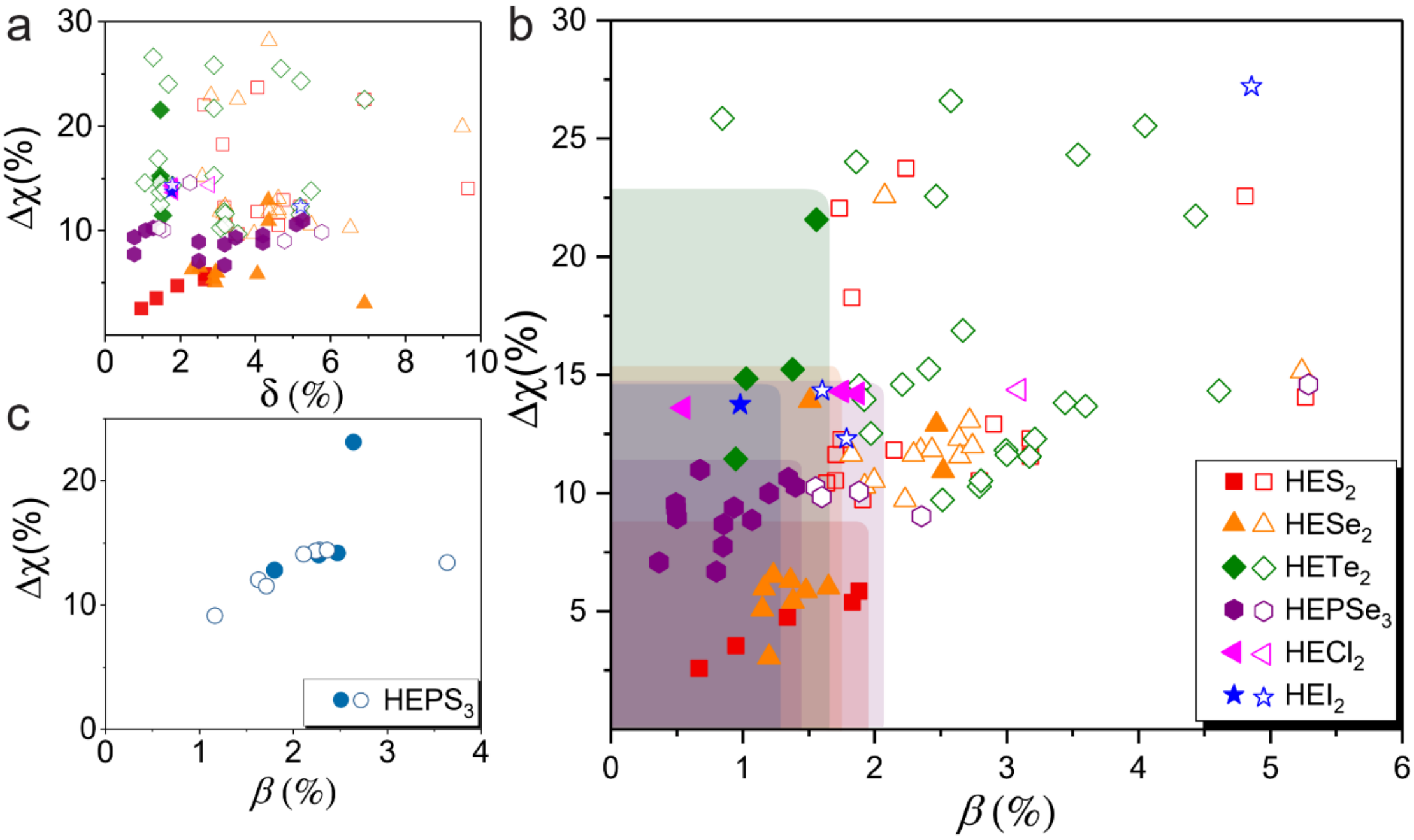

Figure 7. The dependence of phase behavior on the characteristic elemental and structural features in different HEX systems. (a) Conventional Hume-Rothery rule using electronegativity mismatch ($\Delta\chi$) versus atomic size mismatch ($\delta$) plot. The single phase and the phase-separated ones intermingled and the size mismatch failed to be used as a valid criterion. (b) Replot of the phase diagram using the mismatch of the M-X bond ($\beta$) as the abscissa with its definition specified in the text. In practice, these bond lengths are extracted from the ICSD database, and their average value can be automatically calculated using a crystalline 3D visualization program such as VESTA[42]. (c) The exception of the $HEPS_3$ system, lying outside of other HE-systems. Filled and open notations

denote single and multi-phase formation, respectively.

Only the single phase without any impurities, as determined by their XRD patterns, is counted as solid symbols in Fig. 7b. Accordingly, we discovered that the reported Δχ of $(VNbMoW)S_2$, $(NbMoTaW)S_2$, and $(VNbMoTaW)S_2$ are abnormally high (Table 2). As a result, we explored the standard solid-state techniques to synthesize the powder form of all three materials. The atomic ratio powder was heated to 1000 °C, held for 24 hours, and then quenched in water. Their XRD diffraction patterns show a distinct phase separation, suggesting the real composition of these reported phases may be different from the nominal ones[19].

**Table 2. Structure information and physical properties of high-entropy vdW materials.**

| Composition | Space group | Lattice parameter(Å) | Physical behavior | $\beta$ | δ | Δχ | Ref. |
|---|---|---|---|---|---|---|---|
| $(TiVCrNbTa)S_2$ | P-3m1 | a=b=3.3704(2), c=5.57676(9) | Paramagnet, Anderson insulator | 1.88 | 2.65 | 5.85 | [17] |
| $(TiVNbTa)_{0.8}(CrMo)_{0.2}S_2$ | P-3m1 | a=b=3.364(1), c=5.777(1) | Paramagnet | 1.83 | 2.65 | 5.37 | W |
| $Ti_{0.6}(VCrNbTa)_{0.1}S_2$ | P-3m1 | a=b=3.3918(8), c=5.731(1) | Anderson insulator | 1.34 | 1.92 | 4.73 | [17] |
| $Ti_{0.8}(VCrNbTa)_{0.2}S_2$ | P-3m1 | a=b=3.3993(2), c=5.7193(3) | Anderson insulator | 0.95 | 1.37 | 3.54 | [17] |
| $Ti_{0.9}(VCrNbTa)_{0.1}S_2$ | P-3m1 | a=b=3.4057(2), c=5.7171(5) | Anderson insulator | 0.67 | 0.97 | 2.57 | [17] |
| $(VNbMoW)S_2$ | NA | NA | $CO_2$ electroreduction | 1.7 | 3.57 | 33.03 | [19] |
| $(NbMoTaW)S_2$ | NA | NA | $CO_2$ electroreduction | 1.19 | 3.04 | 36.37 | [19] |
| $(VNbMoTaW)S_2$ | NA | NA | $CO_2$ electroreduction | 1.54 | 3.47 | 34.33 | [19] |
| $(TiVCrNbTa)Se_2$ | P-3m1 | a=b=3.5037(5), c=6.0620(6) | Paramagnet, Anderson insulator | 1.48 | 2.65 | 5.85 | [17] |
| $(TiZrHfNb)Se_2$ | P-3m1 | a=b=3.6311(2), c=6.1280(3) | Paramagnet | 2.47 | 4.34 | 12.9 | [17] |
| $(TiVCrNb)_{0.8}(MnFe)_{0.2}Se_2$ | P-3m1 | a=b=3.5133(7), c=6.0920(2) | Spin glass, $T_g$=16 K | 1.51 | 1.71 | 13.9 | W |
| $V_{0.4}(TiCrNbTa)_{0.6}Se_2$ | P-3m1 | a=b=3.474(1), c=6.055(3) | Paramagnet | 1.38 | 2.9 | 5.41 | W |
| $V_{0.6}(TiCrNbTa)_{0.4}Se_2$ | P-3m1 | a=b=3.456(3), c=6.038(2) | Paramagnet | 1.2 | 6.9 | 3.06 | W |
| $Ta_{0.4}(TiVCrNb)_{0.6}Se_2$ | P-3m1 | a=b=3.497(1), c=6.124(2) | Paramagnet | 1.36 | 2.32 | 6.29 | W |
| $Ta_{0.6}(TiVCrNb)_{0.4}Se_2$ | P-3m1 | a=b=3.4934(7), c=6.162(1) | Paramagnet | 1.16 | 2.98 | 5.96 | W |
| $(TiVNbTa)Se_2$ | P-3m1 | a=b=3.4805(1), c=6.1292(3) | NA | 1.15 | 2.94 | 5.07 | [22] |
| $(TiVCrTa)Se_2$ | P-3m1 | a=b=3.48736(5), c=6.1292(3) | Spin glass, $T_g$=6.2 K | 1.23 | 2.53 | 6.5 | [22] |
| $(VCrNbTa)Se_2$ | P-3m1 | a=b=3.4842(3), c=6.1070(6) | NA | 1.65 | 2.94 | 6.02 | [22] |
| (TiVCrNbTa)SSe | P-3m1 | a=b=3.435(1), c=5.945(4) | Spin glass, $T_g$=3.5 K | 1.68 | 2.65 | 5.85 | [17] |
| $(CoAu)_{0.2}(RhIrPdPt)_{0.8}Te_2$ | P-3m1 | a=b=3.9827(1), c=5.2601(2) | Superconductor, $T_c$=4.5 K | 1.37 | 1.47 | 15.2 | [17] |
| $Co_{0.1}(RhIrPdPt)_{0.9}Te_2$ | P-3m1 | a=b=3.9796(1), c=5.2933(1) | Superconductor, $T_c$ = 2.5K | 0.94 | 1.53 | 11.4 | [17] |
| $(CoRhIrPdPt)Te_2$ | P-3m1 | a=b=3.9413(1), c=5.2967(1) | Metal | 1.02 | 1.47 | 14.8 | [17] |
| $(MnFeCoNi)PS_3$ | C2/m | a=5.9321(3),b=10.2737(5), c=6.7061(3) | Antiferromagnet, Spin glass, semiconductor $T_N$=70K, $T_{g1}$=35K, $T_{g2}$=56K | 2.47 | 1.82 | 14.2 | [17] |
| $(ZnMnFeCoNi)PS_3$ | C2/m | a=5.936(1),b=10.297(2), c=6.713(1) | Spin glass, semiconductor $T_g$= 30 K, $T_{kink}$= 120 K | 2.27 | 1.79 | 14 | [17] |
| $(MgMnFeCoNi)PS_3$ | C2/m | a=5.953(1),b=10.372(3), c=6.7394(6) | Multikinks in MT at 8 K, 42 K, 60 K and 120 K. | 2.64 | 3.91 | 23.1 | [17] |
| | | | Semiconductor | | | | |
| (VMnFeCoNi)$PS_3$ | C2/m | a=5.9431(8), b=10.246(2), c=6.7080(7) | Spin glass, semiconductor $T_g$= 37 K, $T_{kink}$ = 150 K | 2.24 | 1.79 | 14.3 | [17] |
| (CoVMnNiZn)$PS_3$ | C2/m | a=5.936, b=10.278, c=6.713 | Hydrogen evolution reaction | 2.3 | 1.47 | 14.3 | [23] |
| $Co_{0.6}$(VMnNiZn)$_{0.4}PS_3$ | C2/m | a=5.918, b=10.251, c=6.694 | Hydrogen evolution reaction | 1.8 | 1.11 | 12.82 | [23] |
| $Fe_{0.7}$(CrMnZn)$_{0.3}PSe_3$ | R-3 | a=b=6.275(1), c=19.873(1) | NA | 1.2 | 1.07 | 10 | W |
| $Mn_{0.1}Fe_{0.8}$(CrZn)$_{0.1}PSe_3$ | R-3 | a=b=6.273(2), c=19.905(5) | NA | 0.93 | 0.78 | 9.4 | W |
| $Mn_{0.1}$ $Fe_{0.6}$ (CrZn)$_{0.3}PSe_3$ | R-3 | a=b=6.296(1), c=19.825(2) | NA | 1.39 | 1.28 | 10.3 | W |
| $Fe_{0.85}$(CrMnZn)$_{0.15}PSe_3$ | R-3 | a=b=6.275(3), c=19.850(2) | NA | 0.85 | 0.78 | 7.74 | W |
| (MnFeCdIn)$PSe_3$ | R-3 | a=b=6.392(1), c=20.024(2) | Superconductor, $T_c$=4.3 K@46 GPa | 1.34 | 5.08 | 10.6 | [18] |
| $Mn_{0.1}Fe_{0.8}$(CdIn)$_{0.1}PSe_3$ | R-3 | a=b=6.2871(3), c=19.88590(9) | Superconductor, $T_c$=6.3 K@30 GPa | 0.85 | 3.18 | 8.7 | [18] |
| $Fe_{0.7}$(MnCdIn)$_{0.3}PSe_3$ | R-3 | a=b=6.3059(1), c=19.8966(4) | Superconductor, $T_c$=6 K@35 GPa | 1.07 | 4.2 | 8.85 | [18] |
| $Fe_{0.85}$(MnCdIn)$_{0.15}PSe_3$ | R-3 | a=b=6.286(2), c=19.902(2) | NA | 0.8 | 3.18 | 6.68 | W |
| (MnFeZnIn)$PSe_3$ | R-3 | a=b=6.323(2), c=19.999(1) | NA | 0.67 | 5.26 | 11 | W |
| $Mn_{0.1}Fe_{0.8}$(ZnIn)$_{0.1}PSe_3$ | R-3 | a=b=6.282(2), c=19.856(2) | NA | 0.5 | 2.49 | 8.95 | W |
| $Fe_{0.7}$(MnZnIn)$_{0.3}PSe_3$ | R-3 | a=b=6.286(1), c=19.975(1) | NA | 0.49 | 3.47 | 9.34 | W |
| $Mn_{0.1}Fe_{0.6}$(ZnIn)$_{0.3}PSe_3$ | R-3 | a=b=6.285(2), c=19.878(1) | NA | 0.49 | 4.2 | 9.58 | W |
| $Fe_{0.85}$(MnZnIn)$_{0.15}PSe_3$ | R-3 | a=b=6.278(1), c=19.841(1) | NA | 0.37 | 2.49 | 7.08 | W |
| (VMnFeCoNi)$Cl_2$ | R-3m | a=b=3.5839(2), c=17.475(2) | Antiferromagnet, Spin glass, insulator, $T_N$= 14.5 K, $T_g$= 9.5 K | 1.75 | 1.79 | 14.3 | [17] |
| (MnFeCoNi)$Cl_2$ | R-3m | a=b=3.5782(5), c=17.467(6) | Antiferromagnet, insulator $T_N$= 15 K | 1.87 | 1.82 | 14.2 | [17] |
| (VMnFeCo)$Cl_2$ | R-3m | a=b=3.6177(2), c=17.507(2) | Spin glass, insulator, $T_g$ = 7 K | 0.55 | 1.82 | 13.6 | [17] |
| (VCrMnFeCo)$I_2$ | P-3m1 | a=b=4.029(5), c=6.734(1) | Spin glass, insulator, $T_g$ = 7 K | 0.98 | 1.77 | 13.7 | [17] |
| (TiVNbMo)$_4C_2T_x$ | NA | NA | NA | 1.86 | 2.93 | 25 | [39] |
| (TVCrMo)$_4C_3T_x$ | NA | NA | NA | 2.21 | 2.53 | 24.2 | [39] |
| (TiVZrNbTa)$_2CT_x$ | NA | NA | Excellent electrochemical performance | 2.75 | 4.61 | 10.5 | [20] |

Notations: NA (not available); W (our own work);

Another benefit of thinking from the structural-unit perspective is that the design of new HEX compounds loosens the need to require the global structural feature comprised of the distinct constituent elements to share the same crystal symmetry. A noticeable example is (TiVCrNb)$_{0.8}$(FeMn)$_{0.2}Se_2$ which could tolerate more than 20% isomerous alloying of Fe and Mn. Generally, $FeSe_2$ (Pmnn) and $MnSe_2$ (Pa-3) do not prefer the P-3m1 symmetry and the maximum doping limit in the simple TMD materials is usually a few percent due to the lattice mismatch (except that Fe doping to $VSe_2$ could reach 20%[43]). According to the database, orthogonal $FeSe_2$ and hexagonal $MnSe_2$ have octahedral $FeSe_6$ or $MnSe_6$ units as the desired building blocks. Further analyzing the bond distance, we find $r_{Fe\text{-}Se}$ = 2.567 Å and $r_{Mn\text{-}Se}$ = 2.564 Å are close to the average value of $r_{M\text{-}Se}$ (M = Ti, V, Cr, Nb) at 2.517 Å, giving $\beta$ of 1.51, located at the left bottom of Fig. 7b.

This observation could further increase the flexibility of composition choice for the design of HEX.

## 3. Specific HEX Systems

### 3.1 HE-dichalcogenides

HE-dichalcogenides include $HES_2$, $HESe_2$, and $HETe_2$, in which the former two systems can be easily grown into single crystals by using iodine as the transport agent. The preparation of their polycrystals and single crystals requires a quenching procedure, otherwise, the sample will spontaneously decompose into multiphases, implying the importance of entropy in producing the single phase. So far, the reported high-entropy sulfides and selenides are all 1T phase (P-3m1) as they are all quenched from or above 1273 K. Wisely choosing the suitable temperature range while avoiding the phase separation may lead to the discovery of other stacking sequences such as 2H and 3R with fundamentally different properties from their 1T counterparts. Very recently, the insertion of multi-elements in between the HE-dichalcogenides layers has been reported. This intercalation strategy can compensate for the entropy deficit of the layered vdW materials compared with their isotropic 3D counterparts [22].

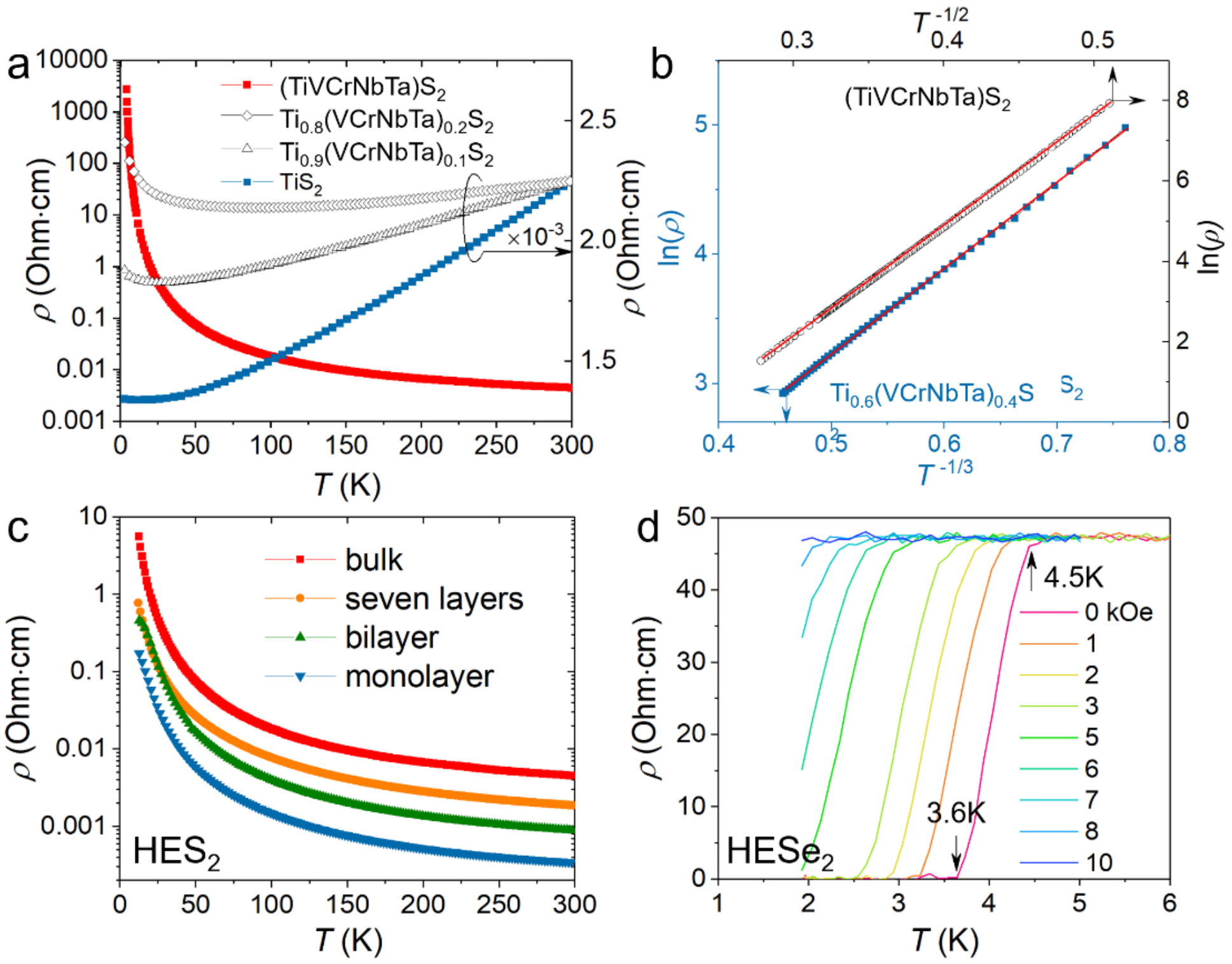

Figure 8. Electrical transport properties of HE-dichalcogenides[17]. (a) Composition-controlled metal-insulator transition of $(TiVCrNbTa)S_2$ with the variation of its entropy content. (b) Two-dimensional variable-range hopping of the resistivity in the $(TiVCrNbTa)S_2$ system. The change of the linear dependence of ln ρ vs T from $T^{-1/3}$ to $T^{-1/2}$ indicates the opening of a Coulomb gap (Efros-Shkovskii gap) at low temperature. (c) Thickness-dependent resistivity of $(TiVCrNbTa)S_2$ from 12 to 300 K. An interesting observation is the anomalous increasing conductivity with decreasing the sample thickness, contrasting to the behavior of the majority of 2D materials. (d) Superconductivity in $(CoAu)_{0.2}(RhIrPdPt)_{0.8}Te_2$. The transition temperatures can be gradually suppressed by increasing external magnetic fields.

The random distribution of the plural elements on the roughly-regular lattice of the HEX family will inevitably modulate the collective transport behavior. As is known, Anderson localization describes a kind of metal-insulator transition (MIT) where an ordered lattice with uncorrelated random potentials would eventually localize the itinerant electrons[44,45]. The concept of HEX is perfectly conformable to the above theoretical framework. $MS_2$ (M = Ti, V, Nb, Ta) are already known to be semimetals or degenerate semiconductors with metallic behavior[46,47]. By imbuing the M sites with high entropy, we could easily achieve MIT in the $(TiVCrNbTa)S_2$ system (Fig. 8a). When the mobility edge extends through the Fermi level, all the charge carriers are localized and migrate in a variable-range hopping manner (VRH, Fig. 8b). The change from a 2D VRH to a $T^{-1/2}$ dependence in $(TiVCrNbTa)S_2$ indicates the opening of a Coulomb gap (known as an Efros−Shklovskii gap[48]). The dependence of electron behavior on the stacking layer number is well affirmed in simple vdW materials. Interestingly, a continuous decrease of the resistivity of $(TiVCrNbTa)S_2$ has been observed with the thinning down of the system to monolayer, which is contrary to the electron transport behavior of other conventional vdW materials (Fig. 8c). The reduction of interlayer scattering may be responsible for this counterintuitive behavior, and a deeper explanation is waiting to be expounded.

Superconductivity is also realized in a range of HEX systems. As a typical example, $(CoAu)_{0.2}(RhIrPdPt)_{0.8}Te_2$ has a superconducting transition temperature ($T_c^{onset}$) at 4.5 K and a zero-resistivity response at 3.6 K (Fig. 8d), comparable to the highest $T_c^{onset}$ in the simple metal-ditelluride systems. It should be emphasized that the maximum $T_c$ values in the previously reported binary or ternary systems usually appear at the boundary of structural instability or on the verge of $Te_2$ dimer breaking, and their superconducting regions are rather narrow, e.g., $Pt_{1-x}Ir_xTe_2$ (~6%)[49] and $Pd_{1-x}Ir_xTe_2$ (~8%)[50]. In contrast, the superconducting state in $HETe_2$ is more robust and easily accessible. A general interpretation of the emergence of superconductivity in metal di-tellurides is the formation of Te-Te dimer. Take $IrTe_2$ for example, the charge transfer from Ir-Ir to Te-Te dimers is deemed as the crucial ingredient to pair the electrons, which is accompanied by the suppression of CDW order. The random atomic occupation will naturally disrupt the charge ordering and introduce abundant fluctuations of Te-Te bonds here and there due to the size dispersion of the plural M-Te bonds. Further enhancement of the $T_c$ is possible by modulating the composition, which directly influences the valence state as well as the strength of electron−phonon coupling. Deliberate incorporation of elements with the local spin state can also give us a platform to investigate the unconventional pairing mechanism of superconductivity and its ostensible competition with magnetism. An apparent suppression of the superconductivity by increasing the doping content of Co has been observed from $Co_{0.1}(RhIrPdPt)_{0.9}Te_2$ to $Co_{0.2}(RhIrPdPt)_{0.8}Te_2$[17].

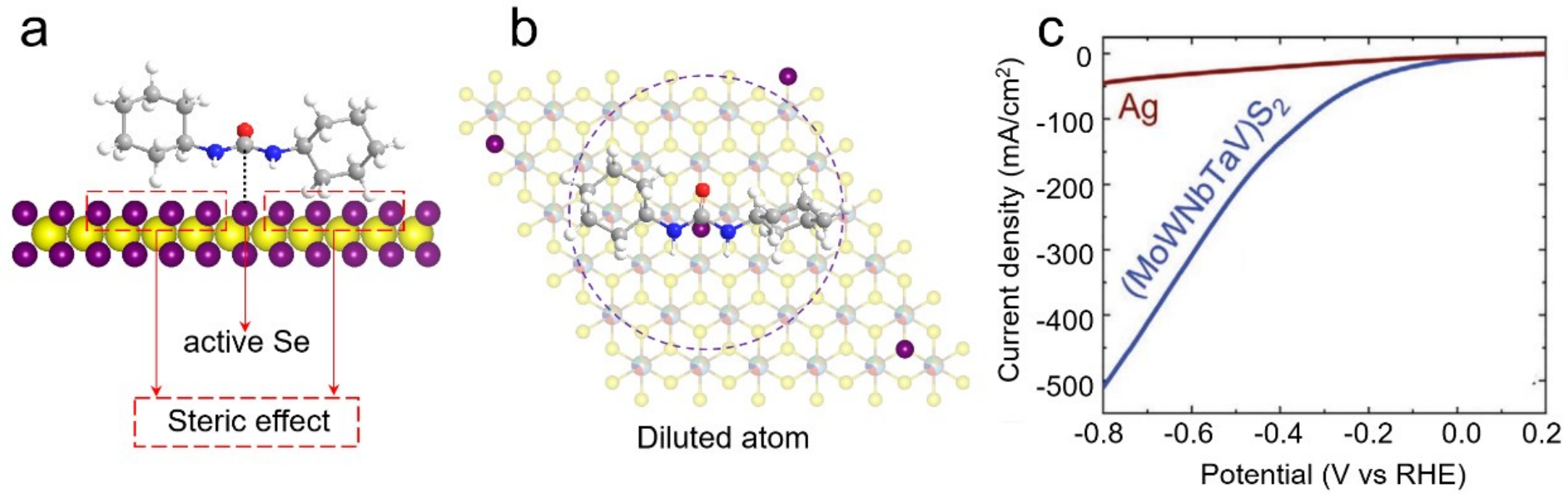

Figure 9. (a, b) Illustration of the concept of isolated atoms in the catalysis by controlling the concentration of expensive or toxic elements to fit the molecular geometry of the desired reaction. (c) The current density of $(MoWVNbTa)S_2$ and Ag nanoparticles plotted against voltage versus RHE using linear sweep voltammetry test mode. (c) is adapted from Ref.[19].

As we know, Se is a good homogeneous catalyst for the reaction of organic ureas. However, the separation of dissolved Se from the product is troublesome. Thus, insoluble heterogeneous catalysts based on Se are much more favorable. The boosted corrosion resistance (demonstrated in Fig. 6) makes $HESe_2$ a promising candidate for the design of highly efficient and durable heterogeneous catalysts. Moreover, further increasing the entropy at the anion site, i.e., the Se site, is a valid way to realize a highly efficient substitution of the expensive or toxic elements. To be specific, for one reaction circle, the catalytic reaction only utilizes one anchoring Se atom, but the large molecule itself will cover a large area (purple circle illustrated in Fig. 9a and 9b), where the surrounding Se atoms do not directly participate in the reaction. By substitution of Se with S, one could design a "diluted atoms" 2D material, $HESe_xS_{2-x}$, for one particular reaction with equivalent catalytic efficiencies.

It is worth pointing out that the negatively charged Se(-2), whether in isolated atoms or not, is quite different from the neutral isolated Se(0). Further incorporation of S with different electronegativity will also alter the chemical potential of Se. However, our recent results (unpublished) show that the negatively charged Se(-2), such as that in $TiSe_2$ and $TaSe_2$, shows a comparable catalytic property to the pure Se, but the dissolving problem still exists. This is where high entropy comes into play, stabilizes the Se, and reduces its consumption. Moreover, the chemical potential changed by incorporating S into Se sites can be balanced by adjusting the alloyed metals. Recent researches on S/Se-based catalysts show much progress, and unique catalytic roles are found such as oxidation of alcohols, olefins, and carbonyl compounds, cyclizations, and ring expansions as reported in reviews[51,52]. Thus, we may expect a unique catalytic activity for S and Se-based HEX based on the synergy effect combined with mixed transition metals.

Recently, Cavin, *et al.* investigated $(MoWVNbTa)S_2$ for the conversion of $CO_2$ to CO[19]. As shown in Fig. 9c, the linear sweep voltammetry tests show the current density for the $HES_2$ at -0.8 V versus reversible hydrogen electrode reaches 0.51 $A \cdot cm^{-2}$, which is more than 10 times higher than that of Ag nanoparticle catalysts. Moreover, the turnover frequencies for CO formation at various voltages hit the highest values among state-of-the-art catalysts reported so far.

### 3.2 HE-phosphorus tri-chalcogenides

Similar to $HES_2$ and $HESe_2$, $HEPS_3$ also shows excellent catalysis performance in $CO_2$-CO conversion reactions. The transition metals M in $MPX_3$ are octahedrally coordinated by six X atoms, which are covalently bonded to P-P dimers, forming an ethane-like $(P_2X_6)^{4-}$ unit. Different from all the other HEX materials, the synthesis of $HEPX_3$ does not require a rapid quenching treatment. Despite that a very slow cooling process (several days at RT) will ultimately lead to phase separation, the conventional furnace cooling process (two hours at RT) yields high-quality monophase.

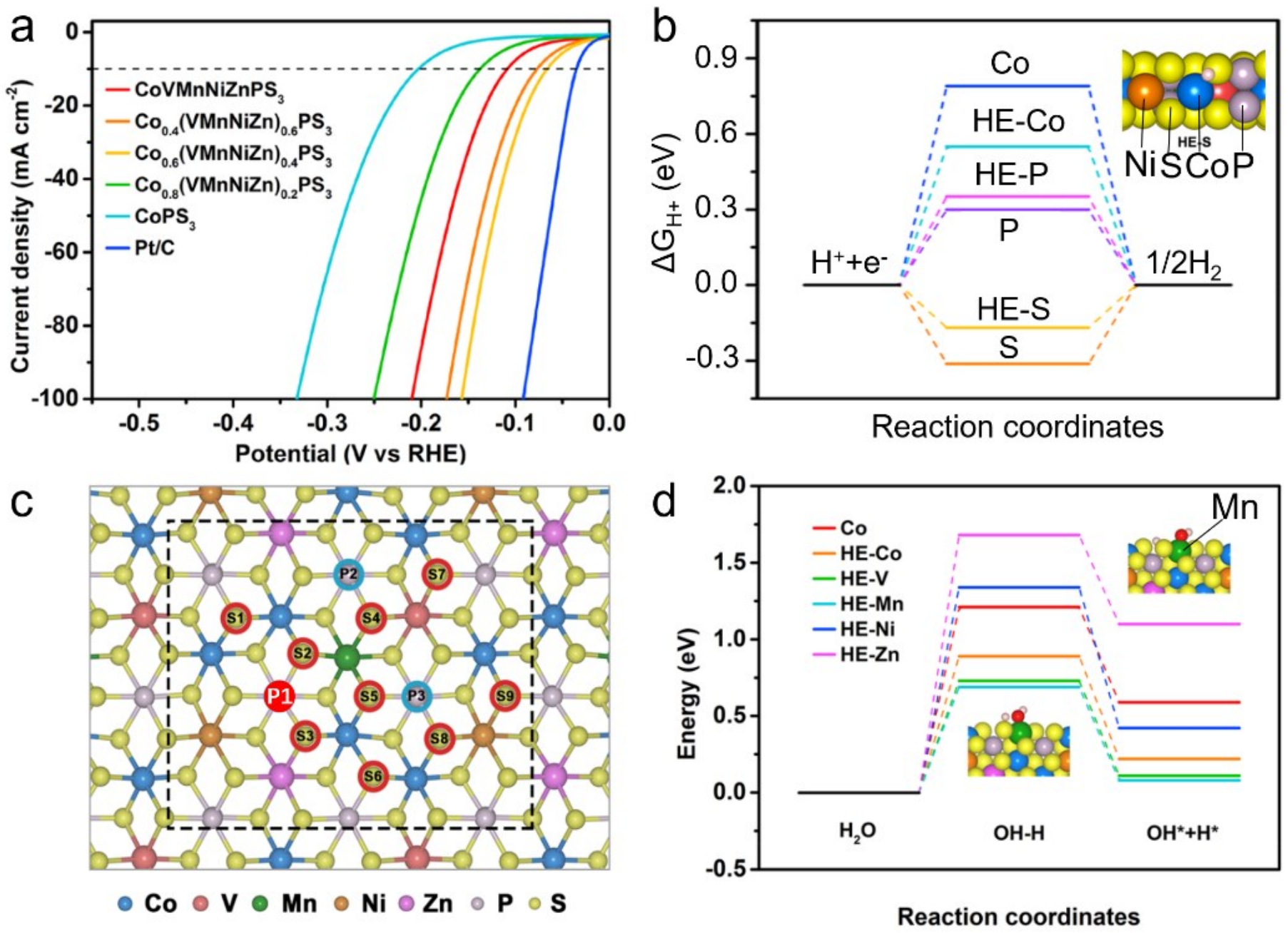

Figure 10. Electrochemical characterizations of $Co_x(VMnNiZn)_{1-x}PS_3$ for HER[23]. (a) Linear sweep voltammetry curves with the variation of entropy. (b) HER free-energy diagram of corresponding edge sites, inset illustrates the adsorption of H on the edge S site (yellow balls). Negative ΔG (HE-S and S) indicates that the adsorption of hydrogen on S is more favored than the desorption process, and vice versa. (c) Basal-plane models of P sites (P1−P3) and S sites (S1−S9) in $Co_{0.6}(VMnNiZn)_{0.4}PS_3$, where the surface P shows the lowest HER free energy. (d) Calculated reaction energy of water dissociation for $Co_{0.6}(VMnNiZn)_{0.4}PS_3$ and $CoPS_3$, including Co, V, Mn, Ni, and Zn sites.

Although noble metals such as Pt, Ru, and Rh have impressive hydrogen evolution reaction (HER) activities, people are looking for more cost-effective and environmentally-benign alternatives. $MPS_3$ system has been extensively studied as a promising candidate with suitable bandgap and high specific surface area. However, theoretical calculations show that only the edge states of $MPS_3$ are active, while the sites on the basal plane are relatively inert to the reaction, which could be due to improper chemical potential. Thus, introducing entropy to $HEPS_3$ with tunable adsorption energies could be a promising technique for improving its catalysis properties. A series of $Co_x(VMnNiZn)_{1-x}PS_3$ are synthesized and their corresponding HER performances are tested[23]. As demonstrated in Fig. 10a, $Co_{0.6}(VMnNiZn)_{0.4}PS_3$ exhibits a much improved HER performance, attaining the greatest record in this family of compounds with an overpotential of 65.9 mV at a current density of 10 $mA\cdot cm^{-2}$. The free energy for hydrogen chemisorption (ΔG) is generally used to understand the active sites of the reaction. The negative value of ΔG indicates that the adsorption of hydrogen is more favored than the desorption process, and vice versa. The favorable reaction will follow the path with the lowest absolute value of ΔG. As shown in Fig. 10b, $|\Delta G_{HE-S}|$ is much smaller than $|\Delta G_S|$, indicating edge site of HE-S is energetically more favorable. Similarly, surface phosphorus is more favored than other atoms (Fig. 10c), and the introduced surface Mn sites boost the water dissociation process (Fig. 10d). Thus, it is the edge-S, on-plane P, and the introduced Mn that collectively contribute to the increased activity.

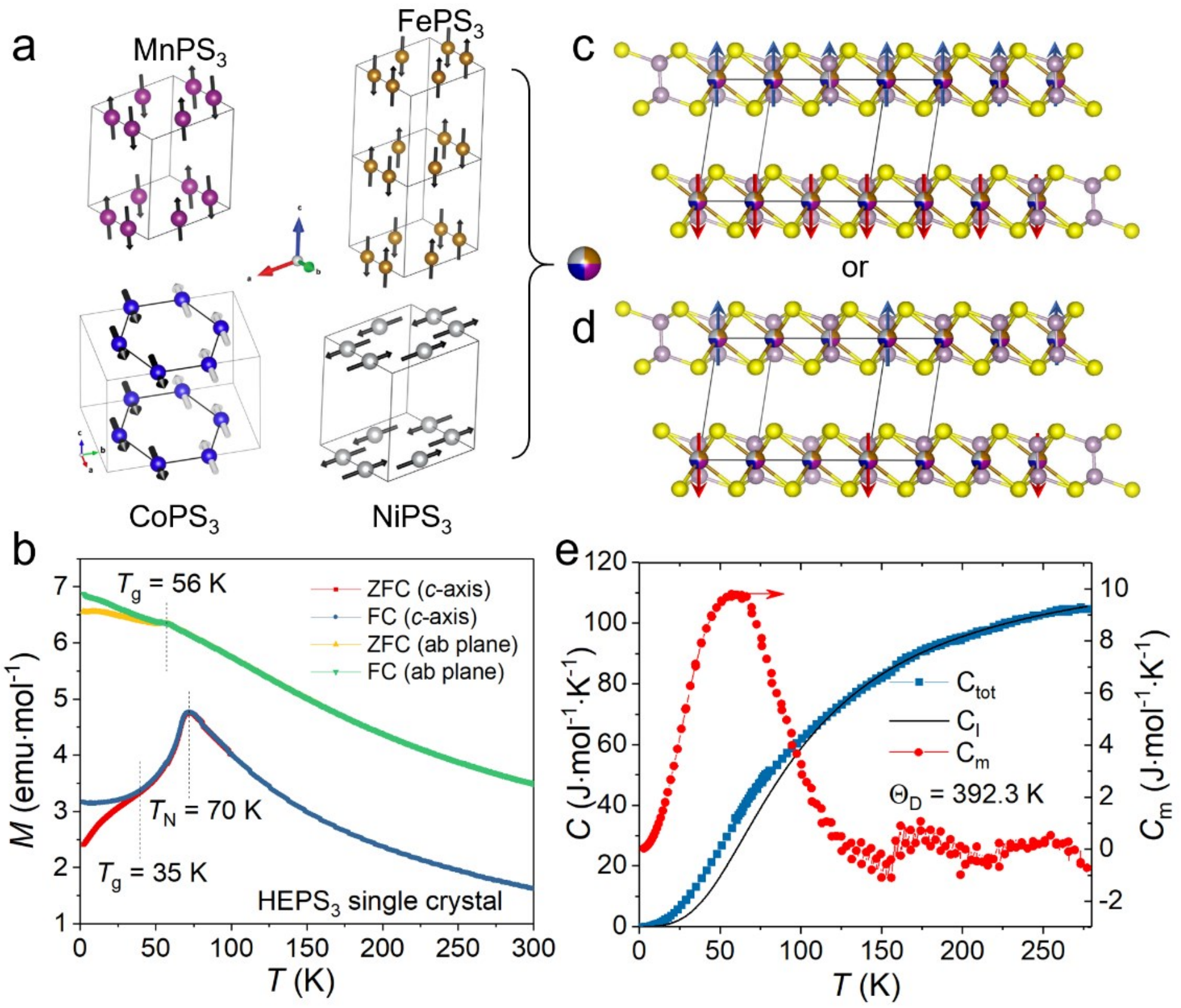

Figure 11. Long-range magnetic ordering in a highly disordered system. (a) Magnetic configurations of individual $MPS_3$ (M = Mn, Fe, Co, Ni). The crystal structure of $HEPS_3$ is identical to $MPS_3$ with the transition metal sites occupied by multiple elements. (b) Magnetization of $(MnFeCoNi)PS_3$ with the external magnetic field H = 500 Oe perpendicular to or within the ab plane. The antiferromagnetic transition ($T_N$) and spin glass transition temperature ($T_g$) are marked[17]. (c, d) Two alternative models of magnetic ordering in the stoichiometric $(MnFeCoNi)PS_3$. (e) Heat capacity of $(MnFeCoNi)PS_3$ from 2 to 280 K. The magnetic contribution (red) is extracted by subtracting the lattice-specific heat (black) by fitting the Debye formula above 160 K.

The simple parent compounds $MPS_3$ have a variety of magnetic properties, i.e., Heisenberg magnets in $MnPS_3$, Ising antiferromagnet in $FePS_3$, and anisotropic (XXZ) Heisenberg magnets in $CoPS_3$ and $NiPS_3$ (Fig. 11a)[53–55]. The compound $(MnFeCoNi)PS_3$ appears antiferromagnetic (AFM) at the intermediate temperature range when cooling down and turns to the spin glass ground state at the base temperature (Fig. 11b). It is thus interesting to ask how the magnetic structure emerges out of the ingredients of different individual magnetic categories (Fig. 11c and 11d). When the magnetic field is measured perpendicular to the *ab*-plane, $(MnFeCoNi)PS_3$ shows a distinct antiferromagnetic transition at $T_N$ = 70 K, which is followed by the emergence of a spin glass transition at 35 K with further decreasing the temperature. When the external field lies in the *ab*-plane, only a spin glass transition shows up at 56 K and the AFM transition becomes almost indistinguishable. Note that the well-defined AFM order emerges in the most random sample. When the composition deviates from the equal molar ratio, as in $Mn_{0.2}Fe_{0.3}Co_{0.25}Ni_{0.25}PS_3$, the AFM transition becomes broad and behaves more like a spin glass. Heat capacity measurement shows a broad hump at around 60 K, indicating the AFM transition to be a Heisenberg type (Fig. 11e).

While the observation of multiple spin glass transitions is understandable, the appearance of an AFM ordering is unexpected. Interestingly, when the atomic ratio of Mn:Fe:Co:Ni deviates from 1:1:1:1, the AFM transition will immediately transform into a spin glass state. This raises a critical question of the origin of a well-defined magnetic ordering arising from a random system and why it becomes more prominent in the most random system.

Another important pending problem to be answered in the random systems is the so-called structural "moderate range order". Due to the 3D nature of most high entropy materials, the structural analysis is difficult to evade the statistical average of the experimental characterization, which inevitably introduces an insurmountable uncertainty and blurs the atomic details[57]. (MnFeCoNi)$PS_3$, as a model system of HEX, offers an ideal playground for atomic mapping (STM, TEM) to clarify the structural features in the short or moderate range owing to easy exfoliation of the material into monolayers. Besides, the interplay of this structural order and the magnetic order in the randomly-distributed regular lattice lends us a fascinating platform to explore the fundamental concepts of statistical thermodynamics, e.g., renormalization, coarse-graining, and scaling.

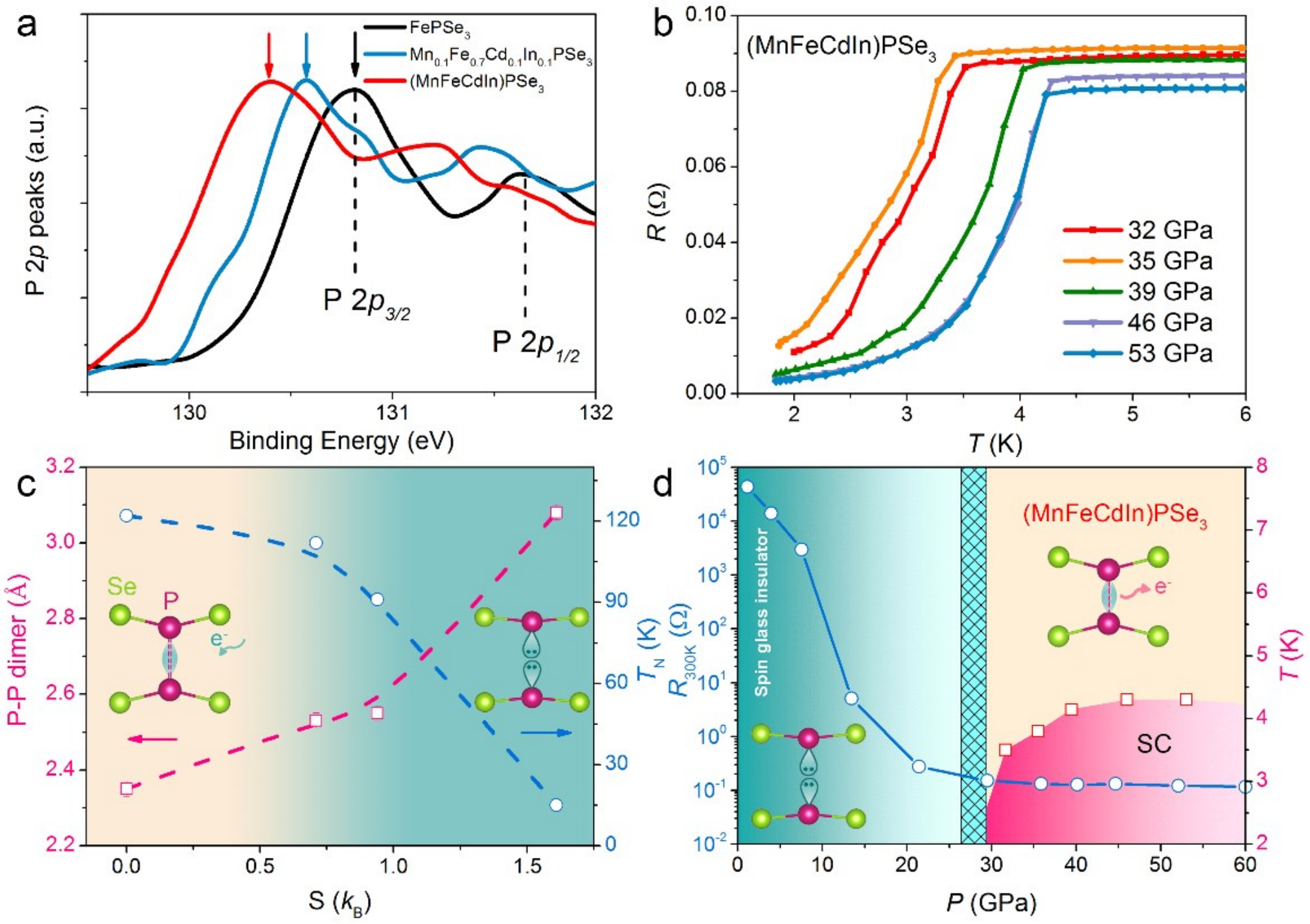

Figure 12. High entropy induced electron donation and superconductivity in the HEPSe3 system[18]. (a) XPS of P 2*p* in (MnFeCdIn)$PSe_3$ compared with that of $FePSe_3$. (b) Metal-superconductor transition of (MnFeCdIn)$PSe_3$ with the external pressure. (c) P-P distance (circles) and Neel temperature (squares) in $MPSe_3$ as a function of entropy. Insets illustrate the breaking of the P-P dimer into two Ps with lone-pair electrons caused by the entropy-driven electron donation. (d) Semiconductor-metal transition and the emergence of superconductivity with the applied external pressure for (MnFeCdIn)$PSe_3$ phase. $R_{300K}$: resistivity at 300 K. The lone-electron pair acts as a charge reservoir that releases electrons under high pressure and is responsible for the out-of-plane collapse.

M in $MPSe_3$ can be flexibly chosen from transition metal ($M^{II}$), alkali ($M^{I}$), alkali-earth ($M^{II}$), and rare-earth metals ($M^{III}$) with an experimental requirement of the averaged valance state is two per

M. This restriction can be found not only in $Fe^{II}PSe_3$, $Ag^{I}In^{III}P_2Se_6$, but also in a variety of nonstoichiometric materials such as $Na_{0.4}{}^{I}Ce_{1.2}{}^{III}PSe_3$ and $Ga_2{}^{III}(PSe3)_3$, where electron doping by introducing trivalent element is compensated by the development of metal vacancies. By counting the electrons of $MPSe_3$, each P shares four electrons with Se to form tetrahedrally covalent bonding and leaves one electron to share with another P to form a P-P dimer. The general P-P distance lies between 1.9~2.1 Å, despite the large variation of the radii of the involved cations.

Superconductivity has long been anticipated in the $FePCh_3$ system (Ch: chalcogen), but so far, it has only been observed in $FePSe_3$ under external pressure with a transition temperature of 4 K. It is suggested that the high-spin to low-spin switch of the Fe $3d^6$ electron-induced magnetic-nonmagnetic transition is crucial for the realization of superconductivity. This explanation is supported by the absence of superconductivity in $MnPSe_3$ ($3d^5$) under pressure but disagreed with the recent neutron diffraction study on the pressurized $FePSe_3$, where the local moment is present at around 5 $\mu_B$ per Fe/Mn atom in different compositions. On the other hand, $FePSe_3$ has the smallest bandgap of 1.3 eV compared with other nonmagnetic counterparts such as $ZnPSe_3$ of 3.2 eV. Electron doping by gating or Li intercalation fails to drive a semiconductor-metal transition in $MPSe_3$ and the underlying mechanism has not been well understood.

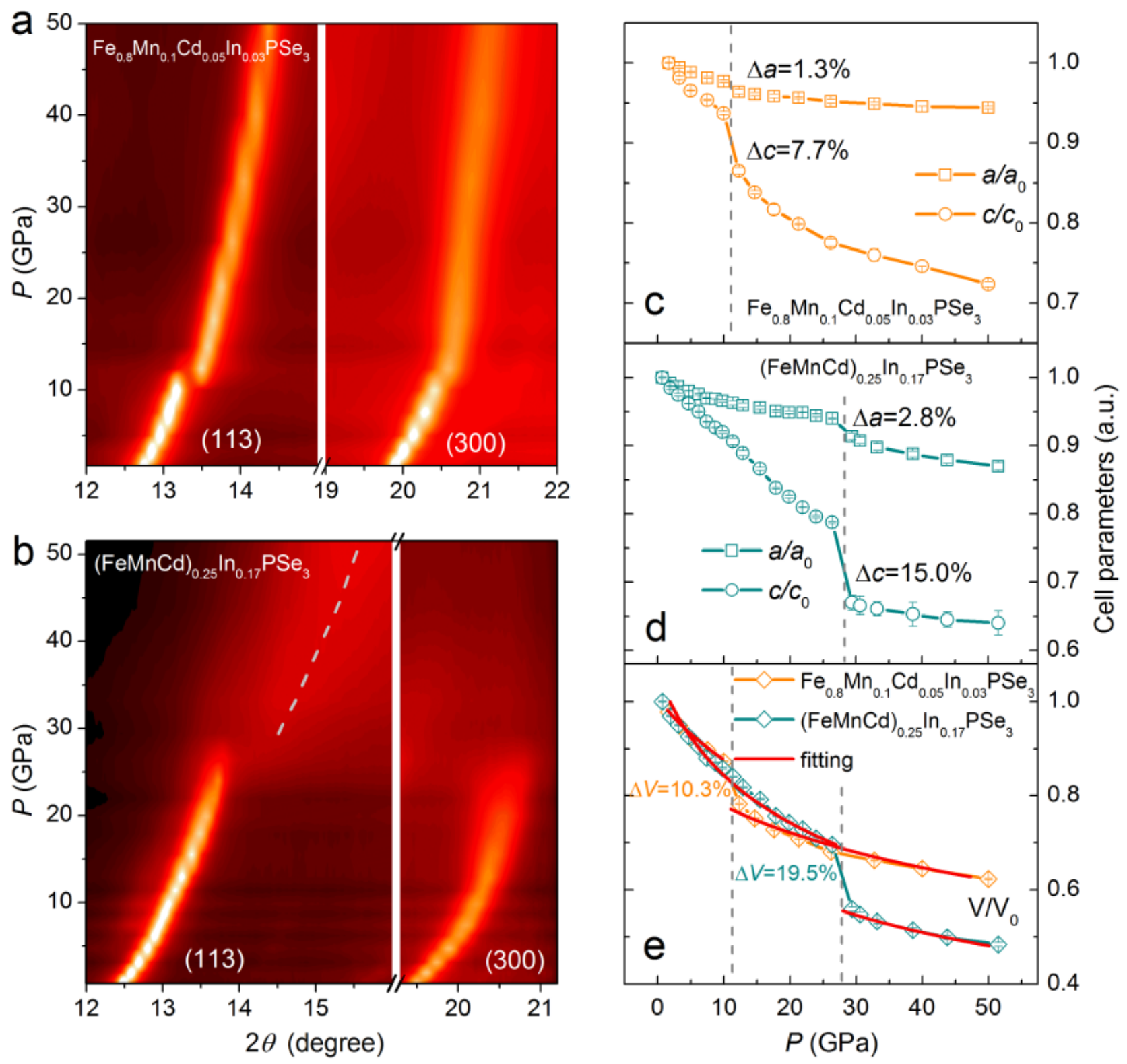

Figure 13. Out-of-plane collapse in pressurized $HEPSe_3$[18]. (a, b) Color contour of the (113) and (300) diffraction peaks for $Fe_{0.8}Mn_{0.1}Cd_{0.05}In_{0.05}PSe_3$ and $(FeMnCd)_{0.25}In_{0.17}PSe_3$ under external physical pressure. (c-e) Pressure-dependent lattice parameters and volume of the unit cell for both samples. The pressure-dependent V of both samples by using the third order Birch-Murnaghan equation of state.

Interestingly, superconductivity can be easily realized in the $MPX_3$ system once entropy comes into play. The randomness introduced by entropy is expected to suppress the magnetic ordering, which should further benefit the emergence of superconductivity. As an additional benefit, we found high entropy could introduce an extra electron donation to the system as evidenced in $In^{3+}$ doped $(MnFeCdIn)PSe_3$. Although the formation of vacancies is inevitable, a prominent discovery is the dramatic enlargement of P-P distance to 3.02 Å (23% expansion) according to our structure refinement[18], while the alloyed samples become more insulating. XPS shows a systematic redshift of the binding energy of phosphorus by increasing the content of In, indicating electron donation has been realized (Fig. 12). The diminished conductivity can be explained by the formation of lone-pair electrons which naturally hamper the conductance of electrons. Similar behavior can be observed in metallic (superconducting) $NaSn_2As_2$ where doping of extra Na will lead to the formation of semiconducting NaSnAs with lone pairs in $Sn^{2+}$. By the same token, we infer the electron doping of $MPX_3$ will also break the P-P dimer, and the generated lone-pair electrons play as charge reservoirs and hinder electronic transportation. On the other hand, as expected, the randomness introduced by high entropy effectively quench the magnetic order. Further applying external pressure, the system went through a subsequent transition from a semiconducting state to a metallic state and ultimately to a superconducting state (Fig. 12b-12d). Noticeably, the superconductivity in $HEPSe_3$ is accompanied by an out-of-plane collapse, different from the in-plane collapse observed in all the individual components (Fig. 13), indicating a different superconducting mechanism from $FePSe_3$. The feasible realization of superconductivity in the $HEPSe_3$ systems highlights the benefit of high entropy in surpassing the restrictions such as electron counting of the $d^6$ rule (the high-spin to low-spin transition leads to a fully occupied $t_{2g}$ orbital in $FeSe_6$ octahedra), and is instructive for the discovery of superconductivity in other systems.

### 3.3 HE-MXene

Not long after the development of high entropy dichalcogenides and phosphorus tri-chalcogenides referred to above, another kind of layered high entropy system, namely high-entropy MXene has been exploited. Different from the synthesis method of HEX, HEXenes were obtained by soft chemical etching of the precursor MAX phases obtained by direct solid-state reactions. The prior acquirement of HE-MAXs is through ball-milling of isomolar alloying metallic powders with certain extra aluminum (10%~30%) and subsequent calcination at 1500 to 1600 °C for several hours followed by a furnace cooling to room temperature. The interlayer aluminum was dissolved by hydrofluoric acid, and the final suspension was filtered and dried to obtain HE-MXene (Figs. 14a, 14b). These soft-chemically deintercalated layer structures present myriad lattice distortions, exposed active sites, and intra-layer buckling which could act as effective heterogeneous nucleation sites of Li atoms for ion battery uses. The resulting mechanical tensions allow the Li nucleation to be effectively guided, resulting in equiaxial crystallite development and avoiding the notorious dendrite growth (Fig. 14c). The achieved excellent cycling stability ranks HEXene as a promising anode material for energy storage.

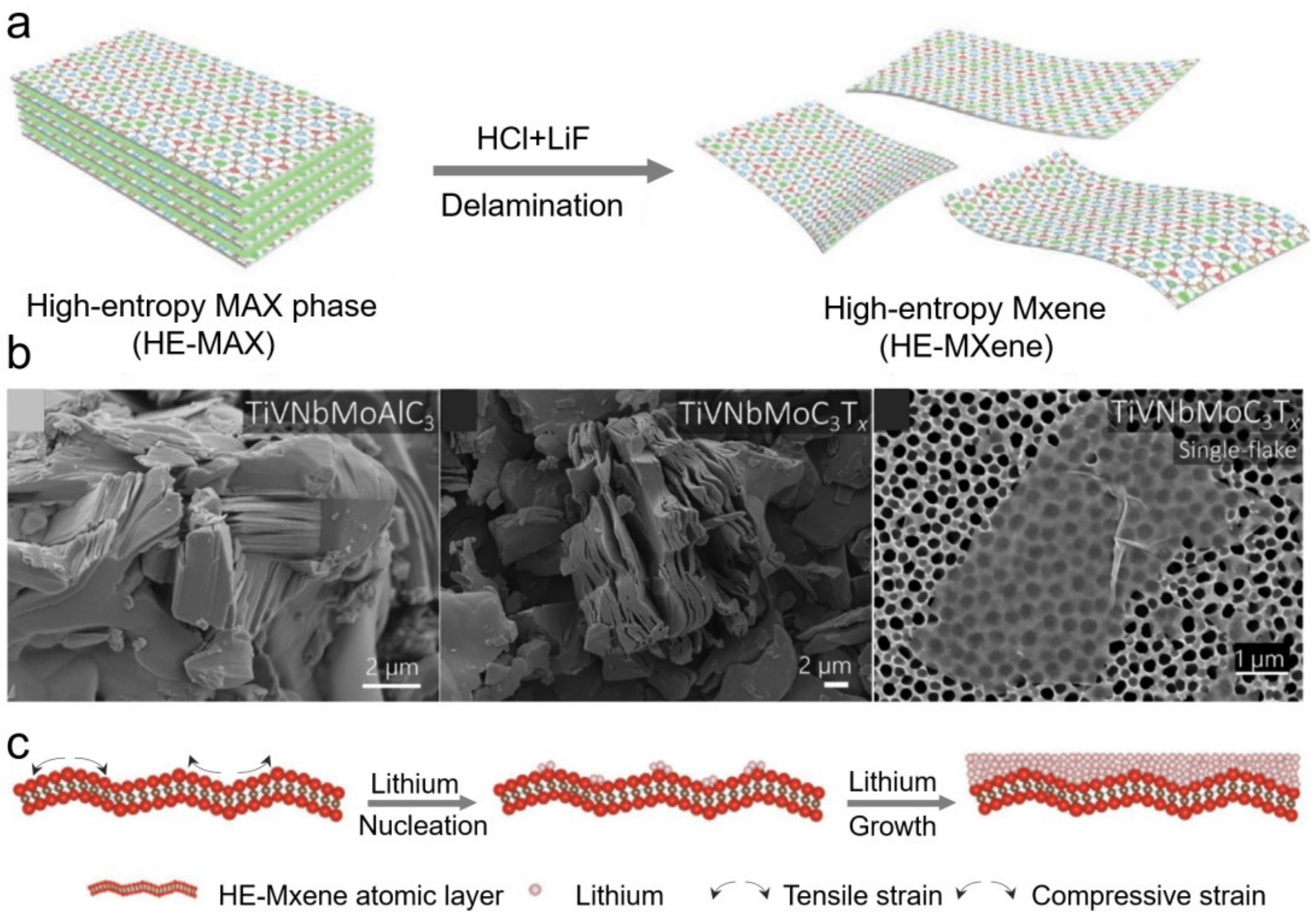

Figure 14. Characterization of high-entropy MXene morphology and structure[20][21]. (a) A schematic illustration of the soft-chemical etching method for obtaining HE-MXene from HE-MAX. (b) SEM micrographs of HE-MAX, HE-MXenes, and a single flake image of $TiVNbMoC_3T_x$ on an alumina substrate. (c) Schematic illustration of the nucleation and growth behavior of lithium guided by strains on HE-MXene atomic layers.

## 4. Concluding remarks: next challenges

The marriage of HEA and vdW materials brings us more degrees of freedom and expands the territory of both (Fig. 15). Only a few characteristics of the brand-new HEX materials have been delineated in this review. To give perspectives and critical assessment of the new-born material subfield, we want to outline the challenges and promises brought about by HEX.

The primary obstacle with high entropy vdW materials is that majority of the discovered physical properties such as superconductivity and MIT have not outperformed their separate components. Although high entropy introduces multiple degrees of freedom, it also comes with disorder, which in principle impairs the performance in general. How to strike a balance between the benefits of high entropy and the drawbacks of randomness is thus one of the key issues. Just like the words by Irnee D'Haenens (who had helped Maiman make the first red ruby laser) describing the laser as "a solution looking for a problem", the potential of HEX is not unearthed yet. There are several yet realized but promising areas waiting to be explored:

(1) Enhanced in-plane Young's modulus: HEA exhibits excellent corrosion resistance and enhanced mechanical properties. The enhanced corrosion resistance in HEX has been substantiated as well[17], while no mechanical properties of HEX have yet been reported. Because of the van der Waals interaction between layers, we anticipate HEX will show an enhanced in-plane bulk modulus inherited from HEA. However, our recent high-pressure measurements reveal the smallest bulk modulus in HEX in both high-pressure and low-pressure ranges (Fig. 13). We are unable to separate the in-plane contribution from that of the out-of-plane since the bulk modulus is retrieved by fitting the volume compression. Future investigations are highly desired to clarify these basic questions. This property, if realized, can be applied to the design of 2D electronic devices with both flexibility and robustness.

(2) Multi-component metal catalyst : Recently research on multi-component catalysts is increasing. Most of them are nanoparticles. Various synthetic routes are reported such as MOF precursor[56]. 2D HEX layer offers a new structural form to work as multi-component catalysts utilizing the cocktail effect of high entropy with tunable carrier density and electron affinity from electronic mixing. Of course, the concept of high entropy can also be applied to anion sites, although the choice of anion is usually limited (S/Se/Te in chalcogens and F/Cl/Br/I for halogens). In Fig. 9a and 9b, we proposed the idea of "isolated atoms" in various catalytic reactions, which may avoid the steric effect, stabilize the alloyed anions, and reduce the consumption of expensive or poisonous elements.

(3) Surface plasmon resonance: Plasma absorption is widely applicable for photonic applications[57]. Surface plasma characteristics strongly depend on the composition of metallic constituents, thus the flexible combination of metal atoms in HEX provides a broad scope to realize surface plasmon resonance that is hardly covered by single elements, e.g., Au (blue-red region) or Cu (red region). Furthermore, these 2D-HEX layered structures can also be used as templates for selective growth of desired nanoparticles. Complex structured metal nanoparticles have been synthesized using controlled nucleation processes[58]. HEX materials with versatile element coordination thus provide another option to regulate the concentration and growth of the absorbed adatoms.

(4) Tunable ionic/electronic/thermal conductivity: The layered structure and weak interlayer interaction of HEX naturally endows the insertion-desertion of ionic species with electroactivity. Through structure and composition design, HEX can act as effective electrode for battery use. Given the disordered distribution of the alloy constituents, and the intrinsic fluctuation of the two-dimensional layered structure, the long-wavelength phonons will be effectively scattered, meanwhile, the mobility of conducting electrons can be maintained if the Coulomb potentials are sufficiently weakened by the screening electrons which may show enhanced thermoelectric properties[59].

(5) Two-dimensional electronic/spintronic devices: Up till now, the multiple degrees of freedom in HEX and their couplings have not thoroughly exploited to realize fascinating device applications. In principle, electronic/spintronic/valleytronic devices can be constructed by lateral or vertical heterojunction through delicate modulation of interface, interlayer and intralayer features. With tunable stacking order, twisting angle, interaction strength and magnetic moments, the electronic correlation and disordered distribution would induce a range of emergent physical behavior. In practice, it is a trade-off between the entropy introduced degrees of freedom and the accompanied unavoidable randomness.

(6) Chemical short-range order and moderate-range order: HEX can also be used as a unique playground to answer several fundamental questions such as magnetic ordering in atomically random systems. Although long-range magnetic ordering in HEA has been found, its origin has not yet been identified. An important pending problem to be answered in the random systems is the so-called chemical short-range order or moderate-range-order[60]. Due to the 3D nature of all the former high entropy materials, the structural analysis is difficult to evade the statistical average of the experimental characterization, which inevitably introduces an insurmountable uncertainty and blurs the atomic details. On the flip side, HEX materials can be feasibly exfoliated to monolayer for atomic mapping, and thus provide an ideal opportunity, and also for the first time, to directly tie the magnetic order to atomic positions with elemental resolution.

Moreover, the random distribution of cations will also localize the valence electrons and reduce the density of the itinerant electrons, thus weakening the screening of the Coulomb interactions between electrons, and making the correlation of electrons more prominent. With the infinite combination and the multi-degrees of freedom available, rich phenomena like heavy fermion and topological effect also await discovery and exploitation.

Another challenge is to predict the phase-formation rule of HEX materials and the link between certain alloying elements and their physical properties. This has been partially solved by introducing an empirical guideline to facilitate the search of single phase in HEX materials. Nevertheless, the boundary of HEX materials is still waiting to be extended. Up till now, the high entropy concept only infiltrated the TMD and MAX systems very recently. It is deserved to modify the broad materials categories with layered nature to bear high entropy features to uncover new physical properties. The concept of high entropy could also be introduced to other systems such as (quasi-) 1D materials and 0D phases.

(7) Bottom-up synthetic process: Thus far, the reported works mainly focused on the physical properties of these vdW materials in the bulk form. Apparently, the really intriguing thing is to go down to flatland to find out what is left and what is emerging. The exfoliation methods employed recently are effective to obtain few-layer or even monolayer HEX. Nevertheless, in order to obtain high-quality large-area monolayer HEX, it is necessary to find new bottom-up synthesis methods. The precise control of stacking layers and the twisting angles are indispensable to exploring the parametric dependence of fascinating behavior. The design of lateral or vertical heterojunction devices also requires the refined modification of boundary and interface qualities. All the existing techniques for 2D materials, such as gating, twisting, stacking, compressing, stretching, and banding could all be naturally employed on HEX thin layers.

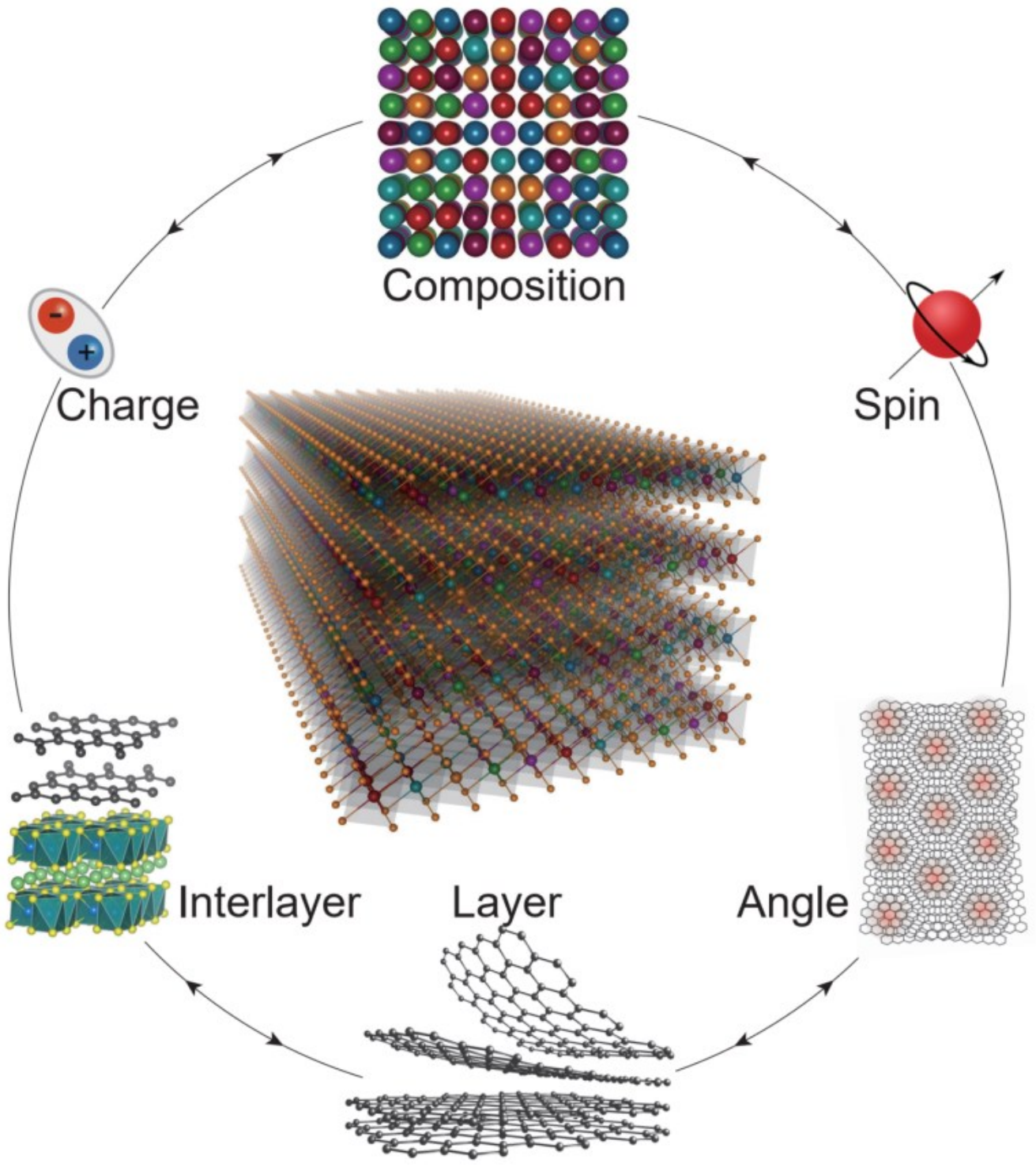

Figure 15. The full set of degrees of freedom in high entropy vdW materials. These freedoms can be employed separately or combined to realize bountiful physical properties or chemical performance[17].

Above all, the concept of high entropy and vdW material has been successfully combined into a new category of materials that could easily be prepared into single crystals, exfoliated into a few layers, and intercalated by different ionic/molecular species. Various physical properties such as superconductivity, magnetic ordering, and metal-insulator transition are demonstrated. The remaining challenges and opportunities are also briefly discussed in this review. Further exploration of this new material family, e.g., its mechanical properties, the origin of long-range magnetic ordering, and band structure of HEX are underway in our lab. Considering its large possibility (multivariate combination and various properties), we think the intricate coupling of their individual property (mentioned above or not) also deserve the focus of the research community, which is beyond the scope of this review but not the promises of this material category. The discovery of HEX materials largely extends the investigation of 2D materials from limited combinations to less limited possibilities.

**Author information**

† These authors contribute equally.

X.L.C.: chenx29@iphy.ac.cn

H.H: hosono@mces.titech.ac.jp

**Acknowledgment**

This work is financially supported by the MoST-Strategic International Cooperation in Science, Technology and Innovation Key Program (No. 2018YFE0202600) and National Key Research and Development Program of China (No. 2021YFA1401800). H.H. was supported by grants from the MEXT Element Strategy Initiative to Form Core Research Center (No. JPMXP0112101001) and JST-Mirai Program (No.JPMJMI21E9).

**TOC:**

**Combining the multiple degrees of freedom inherited from both high entropy systems and van der Waals materials, high entropy van der Waals materials brings about rich emergent physical behavior (superconductivity, thermoelectricity, etc.) and excellent chemical performance (corrosion resistance, heterogenous catalysis, etc.) and is promising for further device applications.**